\documentclass{jfm}
\usepackage{graphicx,amsmath,amssymb,subfigure,psfrag,natbib,multirow,enumerate,overpic}
\usepackage{setspace}
\usepackage{stmaryrd}
\usepackage{epsfig}

\usepackage{datetime}
\usepackage{bm}
\usepackage{siunitx}
\renewcommand{\vec}[1]{\bm{#1}}
\usepackage{epstopdf}
\usepackage{url}

\makeatletter

\newcommand{\Rmnum}[1]{\expandafter\@slowromancap\romannumeral #1@}
\makeatother

\DeclareMathOperator*{\argmin}{argmin}

\shorttitle{LGP control}
\title[Drag reduction of a car model by LGPC]
{Drag reduction of a car model\\
	by linear genetic programming control}

\author[R.~Li and friends]
{
R\ls U\ls I\ls Y\ls I\ls N\ls G\ns  L\ls I $^{1}$
{\thanks{Author to whom correspondence should be addressed: ruiying.li@ensma.fr}}, \ns
B\ls E\ls R\ls N\ls D\ns R.\ns N\ls O\ls A\ls C\ls K$^{2,3}$,\break
L\ls A\ls U\ls R\ls E\ls N\ls T\ns C\ls O\ls R\ls D\ls I\ls E\ls R$^{1}$,\ns
J\ls A\ls C\ls Q\ls U\ls E\ls S\ns B\ls O\ls R\ls \'{E}\ls E$^{1}$,\break
F\ls A\ls B\ls I\ls E\ls N\ns H\ls A\ls R\ls A\ls M\ls B\ls A\ls T$^{4}$, \ns	
E\ls U\ls R\ls I\ls K\ls A\ns K\ls A\ls I\ls S\ls E\ls R$^{5}$ \ns
and\
T\ls H\ls O\ls M\ls A\ls S\ns D\ls U\ls R\ls I\ls E\ls Z$^{6}$\break
}

\affiliation{
$^1$ Institut PPRIME, CNRS -- Universit\'e de Poitiers -- ISAE-ENSMA,\\
F-86962 Futuroscope Chasseneuil, France\\[\affilskip]
$^2$ LIMSI-CNRS, UPR 3251, 
F-91405 Orsay cedex, France\\[\affilskip]
$^3$ Technische Universit\"{a}t Braunschweig, 
D-38108 Braunschweig, Germany\\[\affilskip]
$^4$ PSA Peugeot-Citro\"{e}n, Centre Technique de V\'elizy, F-78943 V\'elizy-Villacoublay, France\\[\affilskip]
$^5$University of Washington, Mechanical Engineering Department, 
Seattle, WA 98195, USA\\[\affilskip]
$^6$Laboratorio de FluidoDin\'{a}mica, CONICET - Universidad de Buenos Aires,\\
Paseo Colon 850, Ciudad Aut\'{o}noma de Buenos Aires, Argentina\\[\affilskip]
}

\pubyear{???}
\volume{???}
\pagerange{???--???}
\date{\textbf{DRAFT\qquad\ddmmyyyydate\today\qquad\currenttime}}
\usepackage[usenames,dvipsnames,svgnames,table]{xcolor}

\definecolor{Gray}{gray}{0.9}
\definecolor{LightCyan}{rgb}{0.88,1,1}

\begin{document}

\maketitle
\begin{abstract}
We investigate open- and closed-loop active control for aerodynamic drag reduction of a car model. Turbulent flow around a blunt-edged Ahmed body is examined at $Re_{H}\approx3\times10^{5}$ based on body height. The actuation is performed with pulsed jets at all trailing edges combined with a Coanda deflection surface. The flow is monitored with pressure sensors distributed at the rear side. We apply a model-free control strategy building on \citet{Dracopoulos1997nca} and \citet{Gautier2015jfm}. 
The optimized control laws comprise periodic forcing, multi-frequency forcing and sensor-based feedback including also time-history information feedback and combination thereof. 
Key enabler is linear genetic programming as simple and efficient framework for multiple inputs (actuators) and multiple outputs (sensors). 
The proposed linear genetic programming control can select the best open- or closed-loop control in an unsupervised manner. Approximately 33\% base pressure recovery associated with 22\% drag reduction is achieved in all considered classes of control laws. Intriguingly, the feedback actuation emulates periodic high-frequency forcing by selecting one pressure sensor in the optimal control law. Our control strategy is, in principle, applicable to all multiple actuators and sensors experiments.
\end{abstract}


\section{Introduction}
\label{ToC:Introduction}

Drag reduction of road vehicles has become a cornerstone challenge due to the increasing need of the reduction of greenhouse gas emissions and corresponding fuel consumption.
Aerodynamic drag represents over $65\%$ of the total power expense \citep{Hucho1998book, McCallen2004aiaa} at highway speeds.
In particular, the low pressure in the wake resulted from the flow separation causes the form drag and constitutes an important portion of the aerodynamic drag for the bluff form vehicles. 
Hence, the manipulation of wake flow provides a great potential to achieve the drag reduction by increasing the base pressure.
Flow control can contribute to fulfil these requirements.

Flow control over bluff bodies can be classified into three groups: passive, active open-loop and active closed-loop controls. Passive control has been widely applied on bluff bodies.  The use of base cavities and boat tails is considered to be one of the most effective devices for base drag reduction \citep{choi2014aerodynamics}. 
Such passive approaches, however, are restricted by design 
and practical considerations and cannot be 'turned off' when not needed.

Facing the constraints of passive control strategy, studies on active flow control (AFC) has rapidly emerged in recent decades. AFC can imitate the effects of passive control. In addition, AFC may be turned on or off depending on the requirement. 
\citet{Cattafesta2011arfm} 
give an extensive overview of possible actuation mechanisms, 
whereas \citet{Choi2008arfm} present the most common AFC approaches on bluff bodies. 
AFC can be performed in an open-loop manner, \textit{i.e.} the control is pre-determined and independent of the flow state. 
The Ahmed body~\citep{Ahmed1984sae} is a widely studied model as its wake depicts many similarities with a road vehicle.
Most of the studies are dedicated to manipulate the wake by forcing the separated shear layer.
These include the application of steady blowing or suction of air flow \citep{Roumeas2009ijnmf,Aubrun2011ef} or unsteady synthetic or pulsed jets \citep{Glezer2005aiaa,Roumeas2009ijnmf,Park2013sae,Joseph2013ef,oxlade2015jfm,seifert2015lab} on the separation trailing edges. 
AFC can be combined with passive deflected surfaces to gain additional base pressure recovery by enhancing the shear layer deflection \citep{Englar2001sae,Englar2004springer,schmidt2015exp,Barros2016jfm}. 

\graphicspath{{./Figures_pdf/}}
\begin{figure}
	\centering
	\includegraphics[width=0.7\textwidth,keepaspectratio]{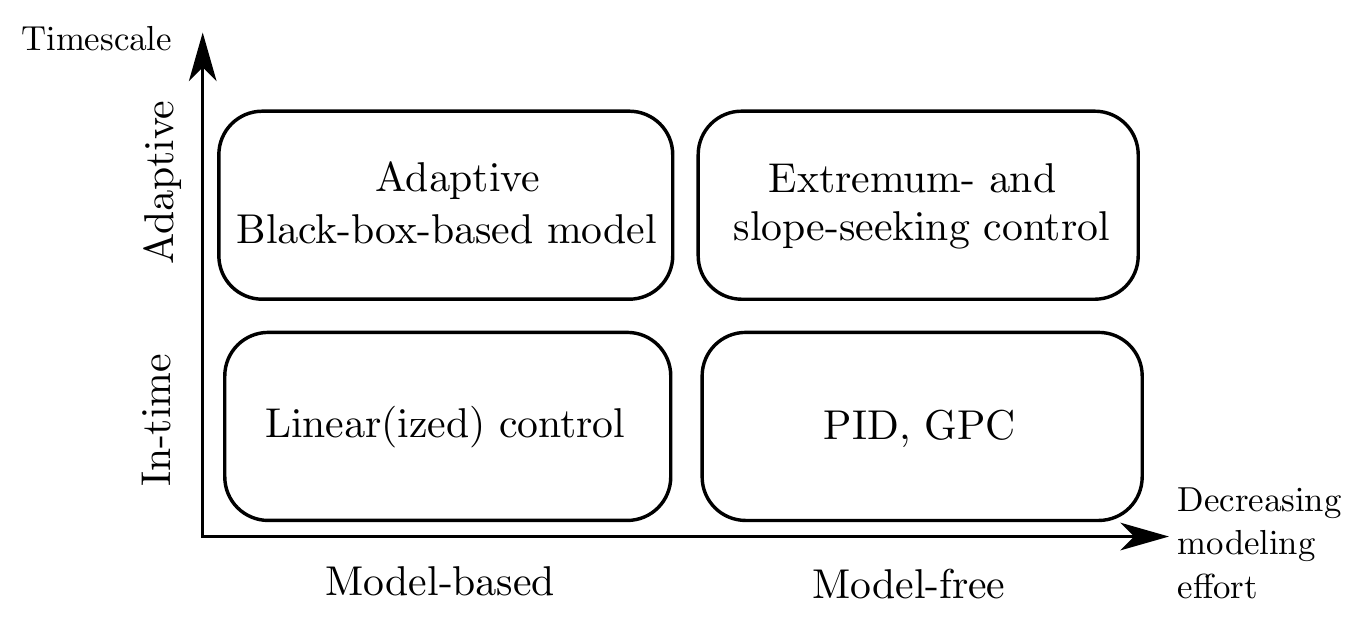}
	\caption{Schematic illustrating popular choices for control design (GPC: Genetic Programming Control).}
	\label{fig:outline_CL_intro}
\end{figure}   
Closed-loop control offers further potential to improve the actuation efficiency by adapting the control to changing flow conditions. 
The actuation is determined by the sensors recording the flow state. 
Depending on the operating timescale of controllers and the required design effort, 
most literature on closed-loop control falls in one of four categories shown in figure \ref{fig:outline_CL_intro}.
There exists a well established framework for the stabilization of laminar flows with in-time model-based control.
'In-time' means that the controller operates on the timescale of the physical processes.
The control may be based on a local linearization of the Navier-Stokes equation. Various configurations have been studied, such as boundary layer flow \citep{liepmann1982active, bagheri2009input}, circular cylinder wake \citep{roussopoulos1993feedback} and open cavity flow \citep{rowley2006linear,samimy2007feedback}.
However, turbulent flow is characterized by broadband frequency dynamics with complex frequency cross-talk. 
The traditional model-based control design mentioned previously is difficult to implement on these flows because the mathematical modeling of the nonlinearities constitutes a great challenge. 
A large portion of the turbulent flow controllers are derived from a reduced-order model, such as Galerkin \citep{gerhard2003model} or vortex models \citep{protas2004linear}, or simple experimentally obtained black-box models \citep{becker2005robust,henning2005drag, henning2007robust,dahan2012feedback}. 
For the latter categories, adaptive concepts are quite promising to maintain performance goals under uncertainties \citep{garwon2005multivariable}.
'Adaptive' means that the controller operates on a timescale much larger than the physical processes.
The response to adaptive control may be adequately modelled by linear or weakly nonlinear dynamics \citep{Pfeiffer2012aiaa}  by averaging over many strongly nonlinear frequency cross-talk mechanisms.

Alternatively, closed-loop control have been designed in a model-free manner, where no underlying model is required. 
Adaptive approaches can be used to find automatically the optimal actuation parameter by a slow feedback of a working open-loop control.
Extremum and slope-seeking control are the most widely used adaptive controllers.
Drag reduction of a bluff body targeting the lowest cost of global energy consumption has been achieved by \citet{beaudoin2006drag} and \citet{Pastoor2008jfm}.
Although this approach is not in-time, the slow feedback has benefits to maintain the performance despite slowly changing environmental conditions.
In-time model-free control may be performed by a PID (Proportional-Integral-Derivative) controller, which is based on a given parametrized control structure \citep{zhang2004closed}.
In more complex configurations with multiple actuators and sensors, no generic simple recipes for the control law can be offered. 
The challenge of the problem lies on the appropriate selection of actuators, sensors and optimization of control laws under a given specific objective.

In this study we perform a model-free open- and closed-loop control for drag reduction of a car model following \citet{Dracopoulos1997nca}.
We employ a recently developed very general model-free control strategy which comprises open-loop, adaptive and in-time control laws.
Departure point is genetic programming control or GPC \citep{Brunton2015amr}.
Here, the closed-loop control design 
is formulated as a regression problem 
in which the feedback law is optimized
with respect to a cost function.
The regression problem is solved with symbolic genetic programming
using the plant (experiment) to evolve the control law.
This model-free control strategy can detect and exploit
nonlinear actuation mechanisms in an unsupervised manner
as evidenced in several shear flow control experiments 
\citep{Gautier2015jfm,Debien2016ef,Parezanovic2016jfm}.

We present the first GPC experiments for bluff-body drag reduction.
Other innovations for flow control experiments include 
(1) the use of \emph{linear} genetic programming \citep{Wahde2008book} as simpler algorithm,
(2) the first application of GPC for several independent actuators,
(3) a very general ansatz for the control law
incorporating sensor-based feedback, multi-frequency forcing and combinations thereof and
(4) the use of filters in GPC.

The manuscript is organized as follows.
The experimental set-up of the generic car model is described in \S~\ref{ToC:Setup}.
In \S~\ref{ToC:LGPC}, the control strategy based on linear genetic programming is proposed. 
Periodic forcing results are presented in \S~\ref{ToC:Periodic forcing} as a reference. 
Open- and closed-loop control using GPC are described in \S~\ref{ToC:Quasi periodic forcing} and \S~\ref{ToC:Feedback control}, respectively. Section \ref{ToC:Conclusions} concludes with a summary and outlook.

\section{Experimental setup}
\label{ToC:Setup}
In this section, the experimental facility is described, 
following an input-output framework appropriate to flow control.
In \S~\ref{sec:Geometry}, the wind tunnel is outlined.
The actuator system and measurements (pressure sensors, drag and velocity measurements) are then detailed in \S~\ref{sec:actuator} and \S~\ref{sec:Measurements}, respectively.
Section \ref{sec:RT_system} presents the real-time system.  

\subsection{Flow configuration and wind tunnel}
\label{sec:Geometry}
\begin{figure}
 	\centering
 	\includegraphics[width=0.9\textwidth,keepaspectratio]{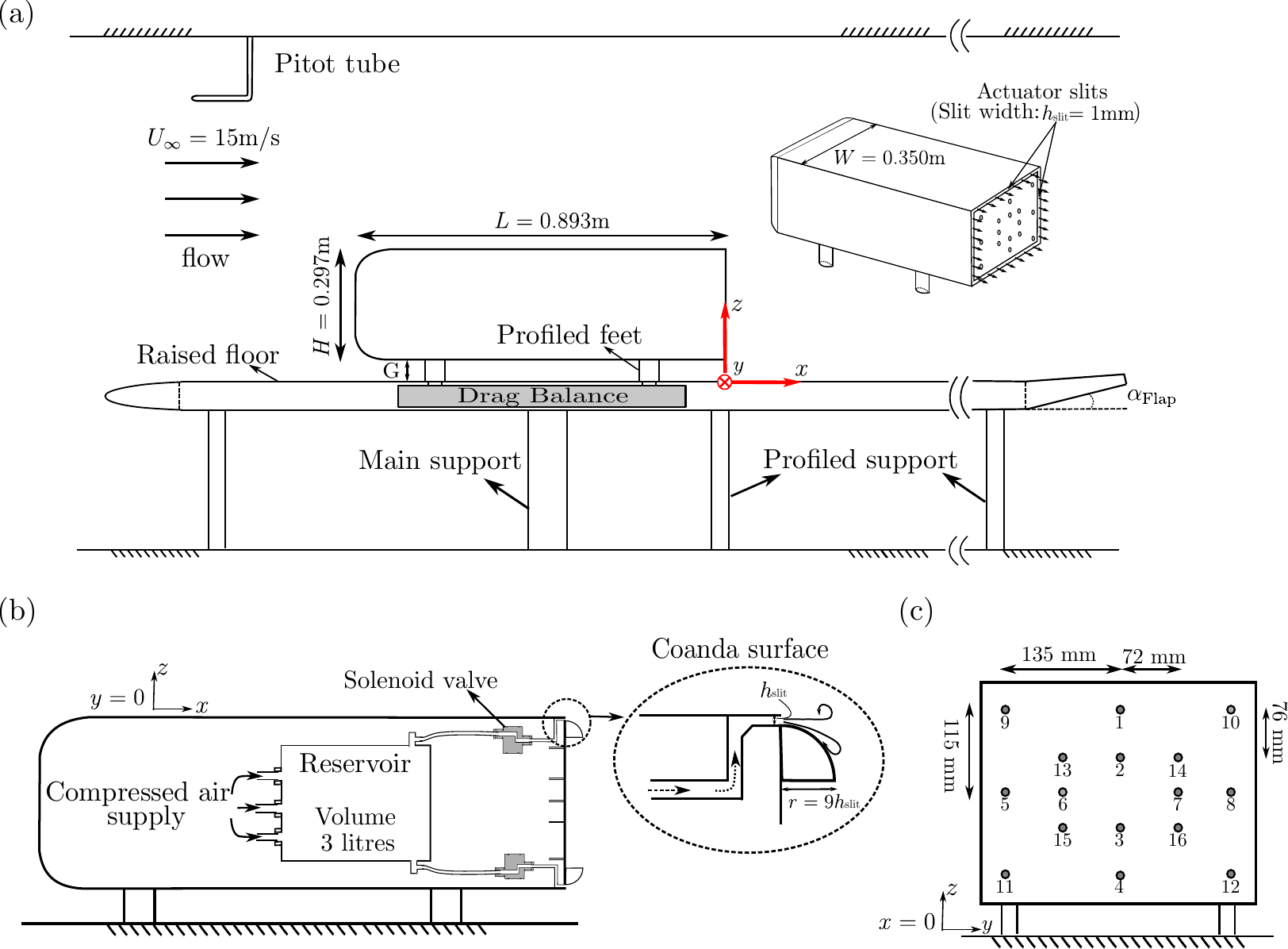}
 	\caption{Experimental setup. (\emph{a}) Wind tunnel and model geometries. Inserted figure shows actuators and sensors on the model.  (\emph{b}) Side view of actuation system. (\emph{c}) Locations of pressure sensors.}
 	\label{fig:sketch_setup}
 \end{figure}

Experiments are conducted in a closed-loop wind tunnel. Its test section is \SI{2.4}{\meter} wide, \SI{2.6}{\meter} high and \SI{6}{\meter} long. The maximum free-stream velocity is about \SI{60}{\meter\per\second} with a turbulence intensity of approximately 0.5\%. A sketch of the test section is presented in figure \ref{fig:sketch_setup}(\emph{a}). The blunt-edged bluff body is a simplified car model similar to the square-back Ahmed body \citep{Ahmed1984sae}. It has the following dimensions: height $H=\SI{0.297}{\meter}$, width $W=\SI{0.350}{\meter}$ and length $L=\SI{0.893}{\meter}$.
$S=HW$ is the frontal area of the bluff body. 
The model is mounted over a raised floor with an elliptical leading-edge to control the boundary layer thickness.
An adjustable trailing edge flap at the end of the raised floor is used to control the incident angle on the leading edge.
Without the model, the zero incident angle is obtained at $\alpha_\text{Flap}=\ang{5.7}$.
After this adjustment, the model is installed with a ground clearance of $G=\SI{0.05}{\meter}$,
as in \citet{Ahmed1984sae}. 
The blockage ratio considering the upper area above the raised floor is 2.2\%.
Additional information on the wind tunnel installation are available in \citet{Barros2016jfm}. 

The flow is described in a Cartesian coordinate system with $x, y, z$ representing streamwise, spanwise (or lateral) and transverse (or normal to ground) directions, respectively. The origin $O$ is placed on the raised floor at the streamwise position of the rear surface.
A Pitot tube mounted on the roof measures the dynamic pressure $q=1/2 \rho U_\infty^2$ from which the upstream velocity $U_\infty$ is deduced. All the results in this paper are obtained with a constant free-stream velocity $U_\infty=\SI{15}{\meter\per\second}$. The Reynolds number based on the height of the model is $Re_{H}=U_\infty H/\nu\approx 3\times10^{5}$, where $\nu$ is the kinematic viscosity of the air. \\  

 \graphicspath{{./Figures_pdf/}}
 
\subsection{Actuator system}
\label{sec:actuator}
The model is equipped with actuator slits at all four trailing edges, as illustrated in figure \ref{fig:sketch_setup}(\emph{a}). The slit width is $h_{\text{slit}}=\SI{1}{\milli\meter}$. The pressured air, which is supplied by a compressed air reservoir, can be blown tangentially to the free-stream velocity through these slits. The reservoir with volume 3 litres is positioned inside the model and connected to the laboratory compressed air network through three \SI{10}{\milli\meter} diameter tubes. The internal pressure of the reservoir is referred as $P_{0}$.

The pulsed blowing is driven by 32 solenoid valves (Matrix\textsuperscript{\textregistered} OX 821.100C2KK) which are installed between the reservoir and the actuator slits, as depicted in figure \ref{fig:sketch_setup}(\emph{b}). These valves are distributed homogeneously along the trailing edges. The zone between the outlet of the valves and the slit exit is specifically designed so that the exiting flow is continuous along the periphery of four edges, as detailed in \citet{Barros2015phd}. The solenoid valve can generate the pulsed jet in ON/OFF mode within the frequency range $[0, 500]\si{\hertz}$. The system enables to control the frequency at the four edges simultaneously or independently.
In the present study, we control only the ON/OFF of the solenoid valves. 
Note that the actuator system has a mechanical time delay between the control command and the fluctuation at the outlet of the slit. This time delay of about \SI{1}{\milli\second} is identified by measuring simultaneously the voltage of the valve and the velocity fluctuation at the outlet of the slit. 
In addition, a rounded surface of radius $9h_{\text{slit}}$ adjacent to each slit exit is installed as an additional passive device in a manner similar to \citet{Barros2016jfm}.
Figure \ref{fig:sketch_setup}(\emph{b}) shows a close-up view of the Coanda surface at the exit zone. 

The actuation amplitude can be characterized by the momentum coefficient:
\begin{equation}
C_{\mu}=\dfrac{S_\text{Jet}\overline{V_\text{Jet}^2}}{SU_\infty^2}
\label{eq:Cmu}
\end{equation}  
where $S_\text{Jet}$ is the slit cross-sectional area, $V_\text{Jet}$ the jet velocity and where the overbar denotes the time average.
The jet velocity is measured at \SI{1}{\milli\meter} downstream the centreline of the slit exit by use of a single hot-wire probe. 
$V_\text{Jet}$ depends on the actuation frequency $f$, duty cycle $DC$ and supply pressure $P_0$. For the closed-loop control, $f$ and $DC$ are unknown before implementing the control law. 
In this study, we choose to maintain a constant initial pressure at $P_{0}^{i}=\SI{4}{\bar}$ before actuation.
When actuation starts, the pressure in the reservoir decreases to about $P_{0}=\SI{1.4}{\bar}$ with a continuous blowing. With a pulsed blowing, $P_0$ depends on the actuation frequency $f$ and duty cycle $DC$.
This initial pressure level is used throughout the experiments except when stated otherwise. The actuation amplitude $C_{\mu}$ is finally obtained by \textit{a posteriori}  measurement of $V_\text{Jet}$ using the registered open- and closed-loop actuation signals. 
 
\subsection{Pressure sensors and measurements}
\label{sec:Measurements}
\subsubsection{Pressure sensors}
\label{sec:Pressure_measurements}
Drag reduction is highly correlated with the base pressure from which the control performance can be quantified. We have 16 pressure taps distributed on the rear surface, as illustrated in figure \ref{fig:sketch_setup}(\emph{a}) with a perspective view. These pressure taps are numbered as presented in figure \ref{fig:sketch_setup}(\emph{c}). The pressure is obtained by differential sensors \textit{Sensortechnics\textsuperscript{\textregistered}} HCLA02X5DB with the following characteristics: operating pressure range $\pm$\SI{250}{\pascal}, response delay \SI{0.5}{\milli\second}, uncertainty due to non-linearity and hysteresis less than 0.25\% of full-scale span. These sensors are connected to the pressure taps through a \SI{0.9}{\meter} long vinyl tube.
The pressure measurements are sampled at a rate of $F_s=\SI{2}{\kilo\hertz}$. The time-history pressure signals will be used as sensor signals in the closed-loop control (see details in \S~\ref{sec:RT_system})
to determine in real-time the actuation. The dimensionless pressure coefficient is defined for each pressure tap $i$ as:
\begin{equation}
C_{p_i}=\dfrac{p_i-p_{a}}{q}, \quad i=1,\ldots,16
\label{eq:Cp}
\end{equation}
where $p_i$ is the measured pressure and $p_{a}$ represents the static pressure of the free-stream.

The tube mounting between the pressure taps and sensors results in distortions between the recorded signals and the pressure values at the taps location.
The recorded signals can be corrected by rebuilding the signals at the  taps location. 
A specially designed coupler having a reference microphone B\&K is applied to obtain a transfer function for each sensor.
An intrinsic impulse response is then derived for each sensor from this transfer function.
The corrected signal is obtained by convolving the impulse response with the measured signal. 
The methodology has been successfully applied in the literature \citep{ruiz2010pressure}. 
The distortion of recorded signals is inferred from the spectrum of the transfer function.
This one presents a low-pass filter behaviour with a linear phase. 
The linear phase leads to a time delay of about \SI{3.5}{\milli\second} (involving the response delay \SI{0.5}{\milli\second} of the sensor) between the fluctuations at the pressure taps and the recorded signals. 
The passband of the low-pass filter, calculated at -3dB in amplitude, is $f\in[0,100]$\si{\hertz} corresponding to a Strouhal number range of $St_{H}=fH/U_{\infty}\in[0,2]$.
This interval covers $St_{H}=0.2$ which is the typical vortex shedding frequency found in the bluff body wake \citep{roshko1955wake}. 
When the flow is forced at frequencies higher than $\SI{100}{\hertz}$, the pulsation strongly affects the wake and consequently the sensing. 
Indeed, the forcing frequency is so energetic that the sensor spectrum still manifests a high energy level at the forcing frequency despite the energy attenuation by the filter. 
Based on this fact, the recorded signals can be used reasonably.
This correction can only be performed \textit{a posteriori} but not on-line. 
Unless the information at the location of pressure taps is needed (the analysis in \S~\ref{sec:QPF control laws}), all the results are obtained directly from the recorded signals without correction.


\subsubsection{Drag measurements}
\label{sec:Drag_measurement}
To quantify the effects of actuation on the drag, the aerodynamic force $F_{D}$ is measured using an in-house unidirectional balance mounted inside the raised floor, as depicted in figure \ref{fig:sketch_setup}(\emph{a}). The principle of the balance is to measure the displacement of two metal plates by use of a 9217A Kistler piezoelectric high sensitive sensor. 
The upper plate is connected to the model through four support profiled feet. The lower plate is fixed to the main support as well as the raised floor. The aerodynamic force on the model creates a downstream displacement of the upper plate against the lower one resulting in an expansion of the sensor. We can then derive the drag force $F_{D}$ from this deformation. The dimensionless drag coefficient $C_{D}$ is defined according to:
\begin{equation}
C_{D}=\dfrac{F_{D}}{qS}.
\label{eq:Cd}
\end{equation}
The pulsed jets create a thrust on the model. 
This is measured in quiescent air  and subtracted from the measured drag at full speed.

\subsubsection{Velocity measurements}
\label{sec:Velocity_measurements}
For analyzing the wake dynamics, velocity fields are obtained using a two-component Particle Image Velocimetry (PIV) system. The measurements are taken in the symmetry plane located at $y=0$. The field dimension spans the whole wake containing entirely the recirculation flow region. The measured region is illuminated by a laser sheet generated by a Nd:YAG laser. Image pairs are captured at a frequency of 
\SI{3}{\hertz} by LaVision Imager LX 16M cameras with resolution of 4920 $\times$ 3280 pixels. The time between a pair of images yielding one velocity field is \SI{90}{\micro\second}. Velocity vectors are processed with an interrogation window of 32 $\times$ 32 pixels with a 50\% overlap, giving a spatial resolution of \SI{2.7}{\milli\metre} corresponding to 0.009$H$. The velocity statistics are computed with 2000 independent images amounting to about $\SI{4}{\minute}$.


\subsection{Real-time system}
\label{sec:RT_system}
For closed-loop control, real-time processing is performed by a Labview Real-Time module, which is implemented on a \textit{National Instrument} PXIe-8820 Real-Time controller running at a sampling rate of $F_{\text{RT}}=\SI{2}{\kilo\hertz}$, where the subscript RT indicates Real-Time. Sensor data acquisition for open- and closed-loop control is performed at the same sampling rate by a \textit{National Instrument} PXIe-6363 DAQ card equipped with 32 analog input channels and 48 digital output channels. Four digital outputs are used to operate the four actuator slits in ON/OFF mode.  Since the solenoid valve cannot respond in less than \SI{1}{\milli\second}, the ON/OFF command needs to have at least \SI{1}{\milli\second}. Under the present sampling rate $F_{\text{RT}}$, this value corresponds to two sampling points. For the effective working of the solenoid valve, a verification is performed before sending the command to the actuators 
to ensure the ON/OFF command lasts at least \SI{1}{\milli\second}.

\begin{figure}
	\centering
	\includegraphics[width=0.6\textwidth,keepaspectratio]{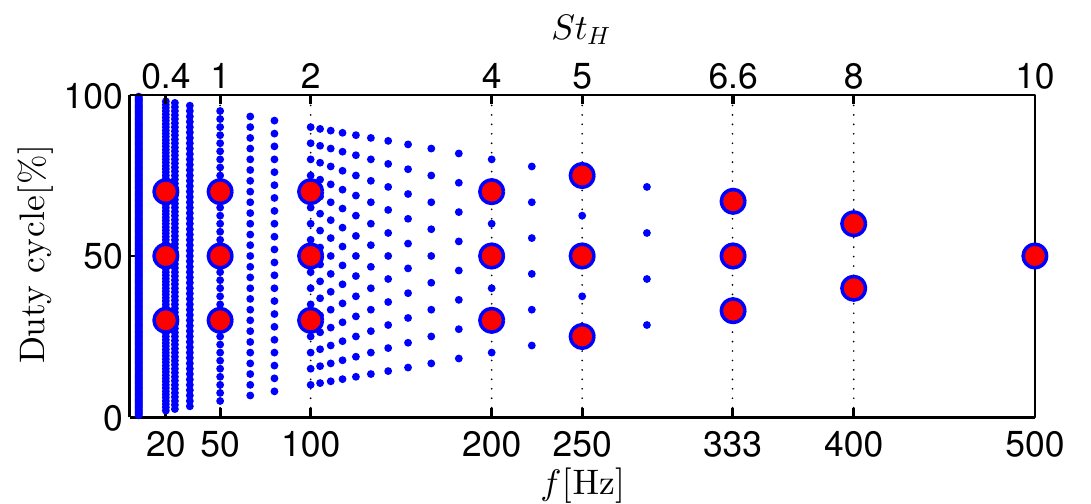}	
    \caption{Ensemble of open-loop frequencies $f$ and duty cycles $DC$ derived from $F_{\text{RT}}=\SI{2}{\kilo\hertz}$. Blue dots represent the combinations of $f$ and $DC$ consistent with $F_{\text{RT}}$. Red dots highlight the cases considered in this study.}
    \label{fig:Fre_DC_2kHz}
	\end{figure}

For a fair comparison of open- and closed-loop control with the same sampling rate $F_{\text{RT}}$, we perform the periodic open-loop forcing using the frequencies and duty cycles derived from the sampling rate $F_{\text{RT}}$. 
Figure \ref{fig:Fre_DC_2kHz} represents with blue dots 
the ensemble of periodic forcing frequencies $f$ and duty cycles $DC$ consistent with the value of $F_{\text{RT}}$. 
The Strouhal number $St_{H}$ is also shown. As the frequency increases, the range of possible duty cycles reduces due to the limited sampling points in one period. The red filled circles highlight the selected periodic forcing cases considered in the following. Hereafter, all the frequencies are given function of $St_{H}$.

%
	\graphicspath{{./Figures_pdf/}}


\section{Linear genetic programming control}
\label{ToC:LGPC}
Following \citet{Duriez2016book},
the control design is formulated as a regression problem:
find the control law which optimizes a given cost function. 
We employ linear genetic programming as powerful and general regression method
for nonlinear functions and for potential multiple extrema of the cost function.
In \S~\ref{sec:Control_problem}, a control problem for drag reduction is formulated. 
In \S~\ref{sec:Ansatz_for_control_function}, 
we introduce a matrix as simple control law representation. 
This law will be evolved with linear genetic programming (LGP) described in \S~\ref{sec:LGP}. 
The experimental realization of LGP is specified in \S~\ref{sec:Param}. 
This evolution is visualized with classical multidimensional scaling method 
as outlined in \S~\ref{sec:Visual}.

\subsection{Control problem}
\label{sec:Control_problem}
The control objective is a net energy saving from drag reduction accounting for the actuation expenditure. 
Both, the thrust-corrected drag reduction
and the actuation energy are determined for the best presented control laws.
However, for the rapid testing of many control laws,
we employ two results of an open-loop study in the same experiment by \citet{Barros2016jfm}.
First, the drag is in good approximation a monotonous function of base-pressure coefficient for all actuation frequencies.
Second, the invested actuation power was found to be a fraction of the drag-related power saving.
In summary, the base-pressure coefficient can be expected to be a good surrogate
for control goal.
The resulting cost function $J$ is defined in terms of the pressure sensors  over the rear side:
\begin{equation}
J=\frac{\langle\overline{C_{p}}\rangle_b}{\langle\overline{C_{p}}\rangle_u} .
\label{eq:J}
\end{equation}
Here, 
$\langle C_{p}\rangle$ 
represents  
the spatially averaged pressure coefficient estimated from $C_{p_i}, i=1,\ldots,16$.
The subscript $b$ indicates the value for the forced flow, 
whereas the subscript $u$  corresponds to the unforced flow.
Thus, the cost function $J$ represents the relative change 
of the spatially and time-averaged base pressure 
by the actuation with respect to the unforced flow. 

The performance of the control law is quantified by $J$.
$C_p$ is negative at the rear side due to the decreased pressure in the wake. 
By definition, $J=1$ for the unforced flow. 
$J<1$ ($J>1$) quantifies the increase (decrease) of base pressure
corresponding to a decrease (increase) of the drag, respectively.
The control task is to minimize the cost function
with a control law $\vec{b}(t)=\vec{K}(\vec{s}(t))$,
where $\vec{b}(t) = \left(b_1(t),...,b_{N_b}(t) \right)^T$ comprises $N_b$ actuation commands
and $\vec{s}(t)=\left( s_1(t),...,s_{N_s}(t) \right)^T$ similarly $N_s$ sensor signals.
In this study, the actuation $\vec{b}$ is performed with pulsed-jets located at the four trailing edges. 
For sensor feedback, 
the argument $\vec{s}$ might be composed of the 16 pressure sensors distributed over the rear surface.
For open-loop optimization, $s_i$ may represent harmonic functions at different frequencies.
The control problem is equivalent to finding $\vec{K}^\text{Opt}$ such that
\begin{equation}
\vec{K}^\text{Opt}(\vec{s})=\mathop{\argmin}_{\vec{K}}{J(\vec{K} (\vec{s}))}.
\end{equation}

\subsection{Ansatz for the control law}
\label{sec:Ansatz_for_control_function}

A control law maps $N_s$ sensor signals 
into $N_b$ actuation commands.
For simplicity, we assume a single-input plant, i.e.~$N_b=1$.
Following linear genetic programming \citep{Wahde2008book},
we assume this control law can be represented by a given maximum number of \emph{instructions}.
These instructions change the content of  $N_r$  \emph{registers}, 
$r_1, \ldots, r_{N_r}$.  
The registers may be variables or constants.
As concrete example, 
we assume that the first $N_s$ registers are initialized with the sensor signals,
the next $N_b=1$ register represents the actuation command, initially zero,
and the next registers contain $N_c$ constants.
These constants are the same for all considered control laws in one optimization.
 
\begin{figure}	
	\includegraphics[scale =1]{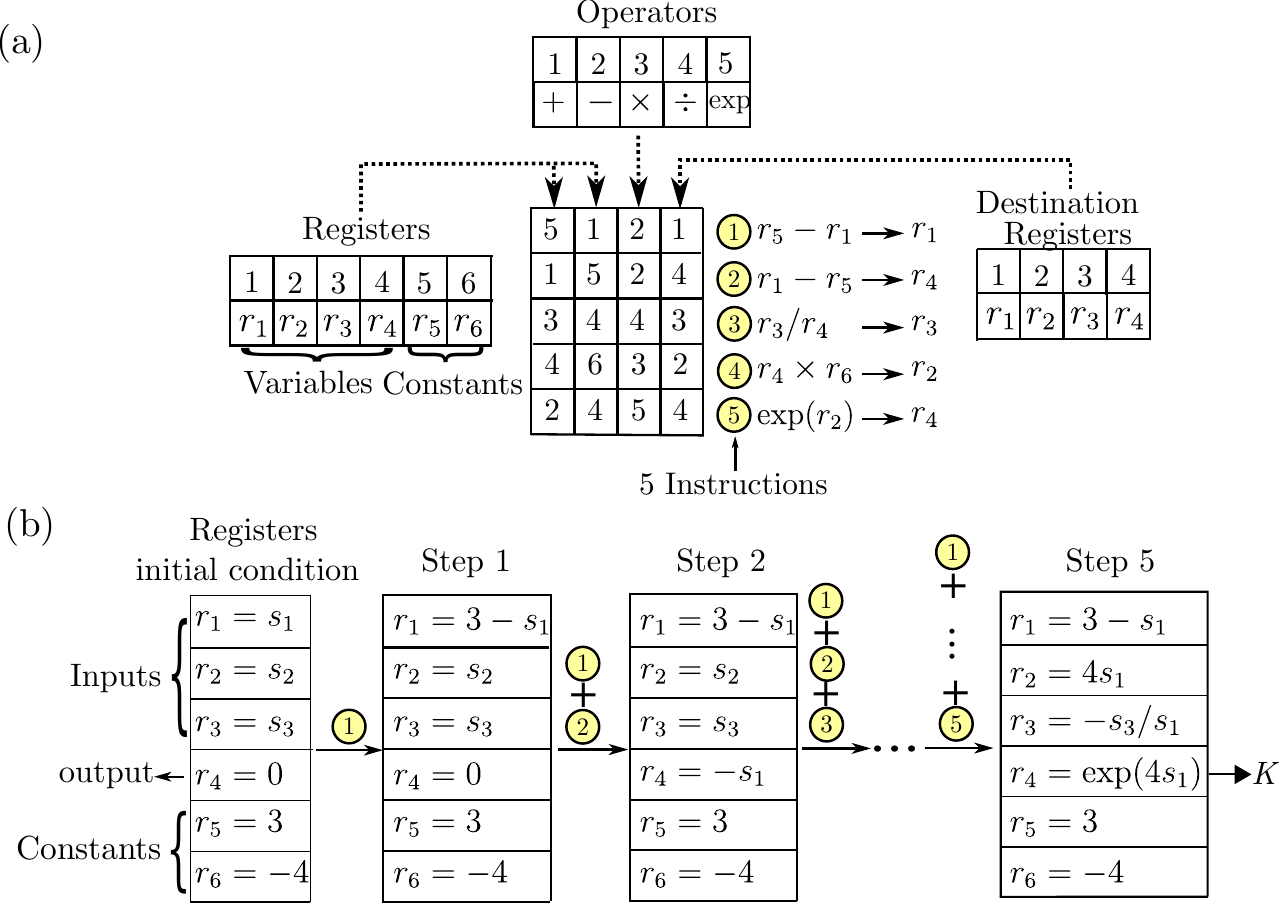}
	\caption{(\emph{a}) An example of matrix $\mathcal{M}$ comprising five instructions ($N_i=5$). 
		The matrix is displayed in the centre of the figure. 
		The five instructions are shown on the right side of the matrix. 
		Let $\mathcal{R}=\{r_1, r_2, r_3, r_4, r_5, r_6\}$  denotes the set of registers, indexed by the integer numbers $\{1,...,6\}$. 
		The first four registers are variables, \textit{i.e.}\ they can be assigned a new value. 
		The last two registers are constants and therefore write-protected. 
		The operand(s) of instructions are coded in the first two columns of the matrix. 
		They can assume any value from $\{1,...,6\}$. 
		The operator set $\mathcal{O}=\{+, -, \times, \div, \exp\}$ is indexed by an integer number $\{1,...,5\}$ and 
		coded in the third column of the matrix. 
		The last column encodes the destination registers, which can be one of the variables from $\{r_1,...,r_4\}$. 
		(\emph{b}) Interpretation of the matrix $\mathcal{M}$. In this example, we have three input registers $\{r_1, r_2, r_3\}$ and one output register $r_4$. 
		Input registers are initialized by the sensors and output register by zero. 
		Step 1 shows the updated registers after implementing the first instruction. 
		Based on this result, we implement the second instruction and obtain the updated result in step 2, etc. 
		The final expressions are obtained after implementing all five instructions. 
		The expression of output register $r_4$ is the targeted function $K$.}
	\label{fig:Figure01}
\end{figure}

An instruction includes an \emph{operation} on one or two registers 
and assigns the result of the operation to a \emph{destination register}, 
\textit{e.g.}, the instruction $r_1:=r_2+r_3$ 
includes two operands, 
the  register $r_2$ and  $r_3$, 
and assigns the result to $r_1$.
One instruction with two operands can be coded as an array of four integers
referring to the two operands, the operator and the destination register, respectively. 
Note that for the instruction with one operand only an array of three integers is required.
However, to maintain a unified representation, 
a fourth integer is equally assigned but ignored.
Consequently, the set of $N_i$ instructions can be coded 
as a matrix $\mathcal{M}$ with dimension $N_{i}\times4$.
An example 
with $N_{i}=5$ is presented in figure \ref{fig:Figure01}. 
Constant registers are write-protected. 
This means that the constants cannot be destination registers 
and their values are initialized at the beginning of a run from a user-defined range. 
One or more variable registers are defined as output register(s). 
The remaining variable registers are referred to as input registers. 
For the decoding, the input registers are initialized 
by the sensor values and the output register(s) by zero. 
The destination registers are updated after each instruction. 
After executing all the instructions, 
the final expression of the output register yields the control law $K$. 
This matrix representation can interpret the instructions efficiently by casting the integer values.

There is only a finite number of control laws 
for a given number of registers $N_r$, 
of operations $N_o$ and of  constants $N_c$:
$$ \left [ N_r \times N_r \times N_o \times (N_r-N_c) \right ]^{N_i} .$$
This number is, however, astromical, even accounting for different matrices
leading to the same control law.
Already the simple matrix of figure \ref{fig:Figure01} 
has over $1.9 \times 10^{14}$ different realizations.
Despite the discrete nature of possible control laws, 
almost any reasonably smooth control law
can be approximated by such a set of instructions
with suitable number of instructions.

Evidently, a combinatorial search of control laws 
and testing in an experiment is not an option.
In contrast, evolutionary algorithms are a near optimum choice. 
In fact, formulating a function from a set of instructions, 
is the constitutive element of Linear Genetic Programming or LGP \citep{brameier2007linear}, as provided in the following section.

\graphicspath{{./Figures_pdf/}}

\subsection{Linear genetic programming}
\label{sec:LGP}

\begin{figure}
 	\centering
 	\includegraphics[scale = 0.9]{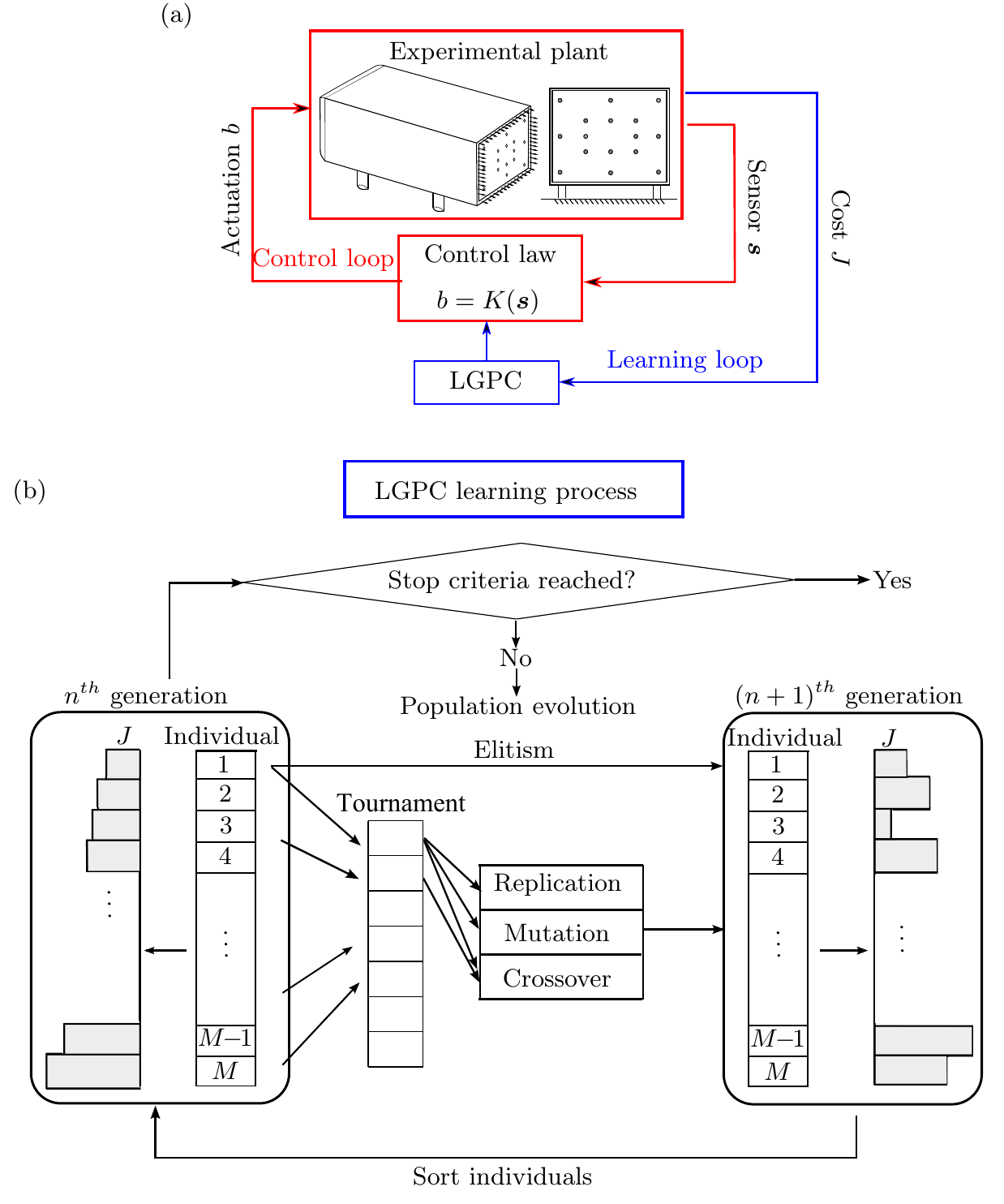}
 	\caption{LGPC implementation on the experiment. 
 		(\emph{a}) Sketch of closed-loop control. 
 		The inner loop performs the real-time closed-loop control. 
 		The plant feeds back the sensor output $\vec{s}$ to the control law. 
 		This control law proposed by LGPC computes the actuation command based on $\vec{s}$ and sends it back to the plant. 
 		A cost $J$ is attributed to the control law after its evaluation during the time $T$. 
 		Then, LGPC in the outer loop uses these costs $J$ to evolve the new population of control laws. 
 		(\emph{b}) LGPC learning process. On the leftmost, an evaluated generation with $M$ individuals is sorted 
 		in ascending order based on $J$. 
 		If the stopping criterion is met, the learning process will be stopped.
 		If not, the next generation (on the rightmost) is evolved by genetic operators (elitism, replication, crossover, and mutation). 
 		After being evaluated, this generation is sorted as indicated by the arrow at the bottom. 
 		We repeat the process from left to right until the stopping criterion is met.}
 	\label{fig:Algorithm}
 	\end{figure}

The employed control optimization has many similarities
with \emph{machine learning control} (MLC) \citep{Duriez2016book}
using the classical tree-based genetic programming (TGP) by \citet{Koza1992book}.
We employ MLC with simpler LGP as regression method 
and refer to it as \emph{linear genetic programming control} (LGPC).
The implementation of LGPC for closed-loop control 
is sketched in figure \ref{fig:Algorithm}(\emph{a}). 
The real-time control process occurs in the inner loop 
with a control law proposed by LGPC. 
The control law is evaluated in the experimental plant during an evaluation time $T$. 
Then, a cost $J$ is attributed to it quantifying the performance of the control law. 
The cost value for each control law is sent to the outer loop where LGPC can learn from them and propose new control law candidates.

\begin{figure}
	\centering
	\includegraphics[scale = 1]{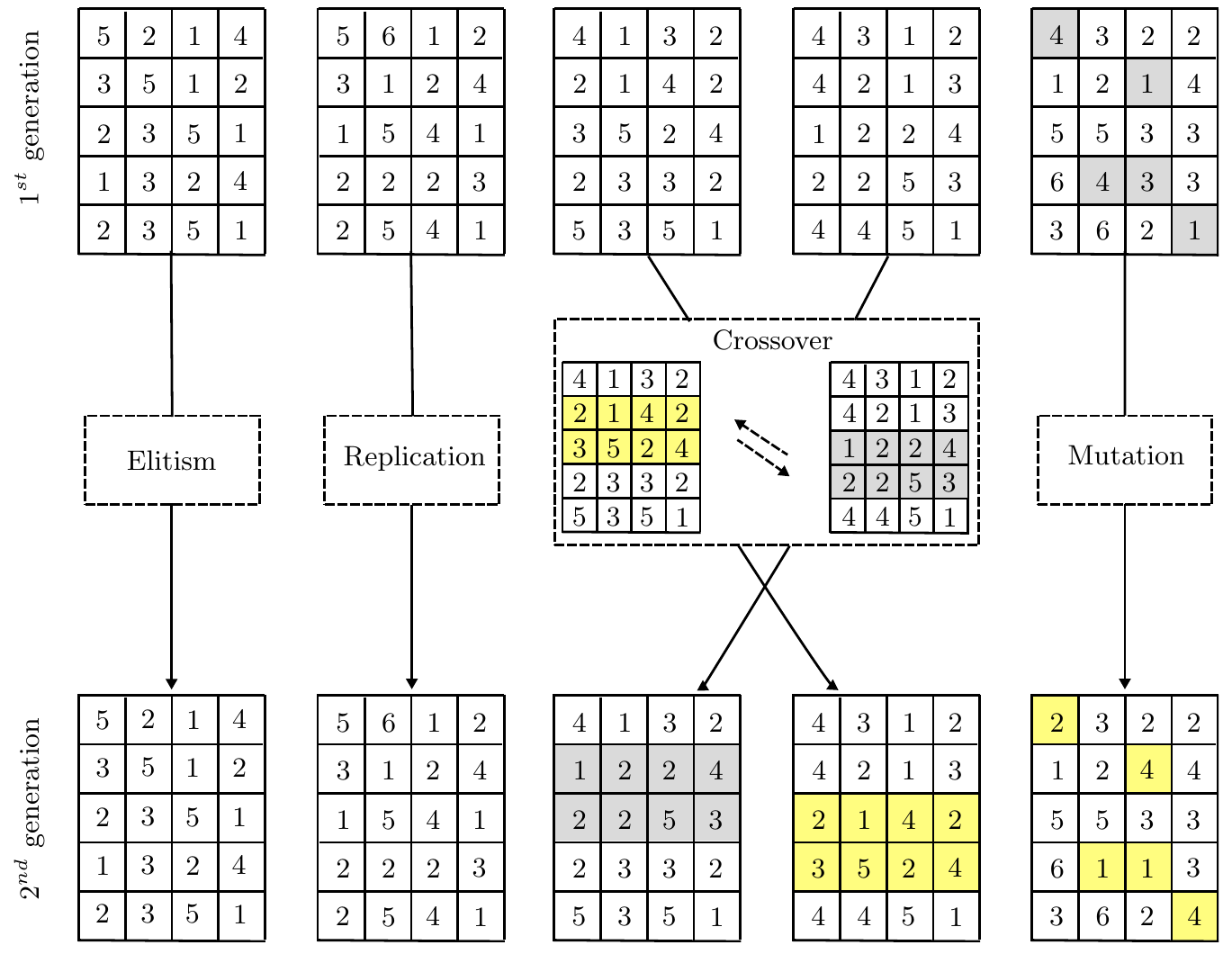}
	\caption{A simple example showing the realization of genetic operations on the individuals for a fixed number of instructions.}
	\label{fig:Figure02}
	\end{figure}

The learning process is detailed in figure \ref{fig:Algorithm}(\emph{b}). 
An initial population of control law candidates, called individuals, 
is generated randomly like in a Monte-Carlo method. 
In the LGPC framework, these individuals are represented by a matrix.
Each individual is evaluated in the inner loop and a cost $J$ is attributed to them. 
After the entire generation is evaluated, its individuals are sorted in ascending order based on $J$. 
The next generation of individuals is then evolved from the previously evaluated one by genetic operators (elitism, replication, crossover, and mutation).
Elitism is a deterministic process 
which copies a given number of top-ranking individuals 
directly to the next generation. 
This ensures that the next generation will not perform worse than the previous one. 
The remaining genetic operations are stochastic in nature 
and have specified selection probabilities. 
The individual(s) used in these genetic operators is (are) selected by a tournament process: $N_t$ randomly chosen individuals compete in a tournament and the winner (based on $J$) is selected.
Replication copies a statistically selected number of individuals to the next generation.
Thus better performing individuals are memorized.
Crossover involves two statistically selected individuals 
and generates a new pair of individuals by exchanging randomly their instructions. 
This operation contributes to breeding better individuals 
by searching the space around well-performing individuals. 
In the mutation operation, random elements  
in the instructions of a statistically selected individual are modified. 
Mutation serves to explore potentially new and better minima of $J$. 
These genetic operations are directly applied to the matrices as depicted in figure \ref{fig:Figure02}. 
After the new generation is filled, the evaluation of this generation can be pursued in the experimental plant. 
This learning process will continue until some stopping criterion is met. 
Different criteria are used.
Ideally, the process is stopped when a known global minimum is obtained (which is unlikely in an experiment).
Alternatively, the termination is triggered by insufficient improvement over the latest generations.
Or a predefined maximum number of generations is reached. 
The targeted optimal control law is the best individual of the last generation.

The range of LGPC is extended by comprising the pressure sensors 
and time-periodic functions into the inputs of the control law. 
It results in a non-autonomous control law $b=K(\vec{s},t)$, where $t$ represents the time. 
The learning process as described previously also applies for this generalized control design.
	\graphicspath{{./Figures_pdf/}}
     
	\graphicspath{{./Figures_pdf/}}

The term \emph{linear} in LGP refers to the linear sequence of instructions,
and not to superposition principle like in differential equations.
The method in itself can provide highly 
non-linear functions as exemplified in figure~\ref{fig:Figure01}.
LGP has two main differences compared to TGP \citep{brameier2005linear}. 
First, multiple usage of register contents results into a graph-based data flow 
which permits a more compact solution than the tree-based structure. 
In addition, the linear instructions are simpler to read and to operate on for MIMO (Multiple-Input Multiple-Output) system 
than the tree-based counterpart. 
Second, in LGP, special noneffective and effective codes coexist. 
The noneffective code refers to the instructions not having an impact on the program output, 
\textit{e.g.} the third instruction $r_3:=r_3/r_4$ in figure \ref{fig:Figure01}(\emph{a}). 
The omission of this instruction will not modify the final output $r_4=\exp(4s_1)$. 
The noneffective code is considered to be beneficial. 
It protects the effective code 
from bad variation effects of genetic operations 
and allows the variations to remain neutral in terms of performance. 
Given these attributes, we choose LGP over TGP to perform this study. 

\subsection{Experimental realization of LGPC}
\label{sec:Param}
LGPC encompasses new features to adapt to experimental applications.
As the solenoid valve works in ON/OFF mode, the output of the control laws is passed through the Heaviside function 
to transform the continuous output to a binary ON/OFF signal, \emph{i.e.} $\mathcal{H}(K(\vec{s}))$, where $K(\vec{s})$ gives a continuous output and $\mathcal{H}$ represents the Heaviside function.
In the following, we assume that $K$ is the binarized control law, \emph{i.e.} $b=1$ and $b=0$ correspond to actuation ON and OFF, respectively.
This binary operation eliminates the amplitude information in the control laws. Therefore, the same actuation signal $b$ can be obtained from different control law expressions. 
The uncertainty in the actuation mechanism may change $J$ 
in different evaluations for the same individual. 
If an individual appears multiple times in several
generations, it is evaluated each time and its cost is the averaged value of all its past evaluations. 
A predetermined number of best individuals in each generation are re-evaluated several times to ensure good and 
robust performance.

The LGPC parameters for this study are displayed in table \ref{tab:LGP_param}. 
\begin{table}
	\centering	
	\begin{tabular}{lr}
		Parameters  & Value  \\	\hline
		Population size & $M=$50 \\
		Tournament size & $N_t$=7 \\
		Elitism & $N_e=1$ \\ 
		Replication & $P_r=10\%$ \\
		Crossover & $P_c=50\%$ \\
		Mutation & $P_m=40\%$ \\ 
		Min. instruction number & 5 \\
		Max. instruction number & 30 \\
		Operations & $+,-,\times, \div, \sin, \cos, \tanh, \log$\\
		Number of constants & $N_c=6$ \\
		Constant range & $[-1,1]$ \\		
	\end{tabular}
	\caption{LGPC parameters in the experiments.}
	\label{tab:LGP_param}
\end{table}
Each generation is composed of $M=50$ individuals. 
An optional pre-selection of individuals is performed for all the generations.
After generating the individuals, each is pre-evaluated based on the pressure signal of the unforced flow.
The resulting actuation command is an indicator for their feedback control performance.
If no actuation ($b=0, \forall t$) or continuous blowing ($b=1, \forall t$) is obtained in the pre-evaluation, 
this individual may be considered a prospectively bad performer and is discarded for evaluation by assigning a high cost value to it.
This pre-evaluation step promotes a fast convergence.

Elitism is set to $N_e=1$, \emph{i.e.} the best individual of a generation is copied to the next.  
The replication, crossover and mutation probability are 10\%, 50\% and 40\%, respectively. 
The individuals on which these genetic operations are performed come from a tournament selection of size $N_t=7$. 
The instruction number varies between 5 to 30 (except where noted otherwise) with a Gaussian distribution. 
Elementary operations comprise $+,-,\times, \div, \sin, \cos, \tanh$ and $ \log_{10}$. 
The operation $\log_{10}$ is protected, \textit{i.e.} $\log_{10}(x)$ is modified to $\log_{10}(|x|)$ where $x$ is the variable. 
If the actuation command at time $t_{k}$ is not a number (NaN) or infinity (Inf) due to the sensitive operator $\div$, 
it is modified to take the command one step before, \emph{i.e.} $b(t_k)=b(t_{k-1})$. 
In addition, we choose six random constants in the range $[-1,1]$.

The evaluation of every individual takes $T=\SI{10}{\second}$. 
This value corresponds to 500 convective time units defined by $H/U_{\infty}$. 
According to the results presented in~\citet{Barros2015phd}, 
the base pressure has converged in this time interval. 
There is a time gap of about \SI{6}{\second} between two individuals for data recording, 
reservoir refilling and communication between LGPC and the control module. 
The best five individuals of any generation are re-evaluated five times. 
Overall, approximately five generations each consisting of 50 individuals are evaluated in less than two hours.

\subsection{Visualization of control laws}
\label{sec:Visual}
LGPC systematically explores the control law space by generating and evaluating a large number of control laws
from one generation to the next.
An assessment of the similarity of control laws 
gives additional insights into their diversity and convergence to optimal control laws, 
\textit{i.e.} into the explorative and exploitative nature of LGPC.
For that purpose, we rely on multidimensional scaling or MDS \citep{Mardia1979book}, a method classically used 
to visualize abstract data in a low-dimensional space.
MDS comprises a collection of algorithms
to detect a meaningful low-dimensional embedding given a dissimilarity matrix
with the purpose of visualizing the (dis)similarity of objects or observations. 
Here, we employ classical multidimensional scaling (CMDS) which originated from the works 
of \cite{Schoenberg} and \citet{YoungHouseholder}.
Let us define $N$ the number of objects to visualize, and $\vec{\mathsfbi{D}}=\left(\mathsfi{D}_{lm}\right)_{1\le l,m \le N}$ a 
given distance matrix of the original high-dimensional data.
The aim of CMDS is to find a centred representation of points 
$\vec{\Gamma} = [\vec{\gamma}^1\quad \vec{\gamma}^2\quad \ldots\quad\vec{\gamma}^N]$
with $\vec{\gamma}^1,\ldots,\vec{\gamma}^N\in\mathbb{R}^r$,
where $r$ is typically chosen to be $2$ or $3$ for visualization purposes, 
such that the pairwise distances of the points approximate the true distances, 
\textit{i.e.} $\vert\vert \vec{\gamma}^l-\vec{\gamma}^m\vert\vert_2 \approx \mathsfi{D}_{lm}$. 
The details of the implementation are given in \S~\ref{Sec:AppA:CMDS}. 
In our case, we have $N=M\times G$ where $M$ is the number of individuals in a generation, and $G$ is the total number of generations.

For measuring the dissimilarity between two control inputs $b^l$ and $b^m$ with $l,m\in\{1,\cdots,N\}$, we define the square of the cross-generational distance matrix $\boldsymbol{\mathsfbi{D}}^2$ as
\begin{equation}
 \mathsfi{D}_{lm}^2 =   \left \langle
                \left\vert b^l \left(\vec{s}\right)
                         - b^m \left(\vec{s}\right)
                \right\vert^2 
                \right \rangle_{l,m} 
+ \alpha\,\vert J^l-J^m\vert.
\label{Eq:DistanceMatrixControlLaws}
\end{equation}
The first term represents
the difference between the $l$th and $m$th control law
averaged over the sensor readings of both actuated dynamics.
Thus, the averaging takes into account the frequency 
and relevance of the sensor reading.
The second term penalizes difference of their achieved costs $J$ with coefficient $\alpha$.
This penalization is important as even very similar actuation time series may be very different in their respective performance.
The penalization coefficient $\alpha$ is chosen as the ratio 
between the maximum difference of two control laws 
(first term of \eqref{Eq:DistanceMatrixControlLaws})
and the maximum difference of cost function 
(second term of \eqref{Eq:DistanceMatrixControlLaws}).
Thus, the dissimilarities between control laws and between the cost functions 
have comparable weights in the distance matrix $\mathsfi{D}_{lm}$.

The first term of \eqref{Eq:DistanceMatrixControlLaws}
can easily be computed. 
Let $Q$ be the number of sensor signals 
recorded with constant sampling frequency
at times $t_q$, $q=1,\ldots,Q$,
both, for actuation under the $l$th and the $m$th control law.
The corresponding sensor readings are denoted by
 $\vec{s}^l(t)$ and $\vec{s}^m(t) $, respectively.
Then, 
the ensemble average $\langle \quad \rangle_{l,m}$ 
of \eqref{Eq:DistanceMatrixControlLaws} is approximated by
$$
\begin{aligned}
\left \langle
                \left\vert b^l \left(\vec{s}\right)
                         - b^m \left(\vec{s}\right)
                \right\vert^2 
                \right \rangle_{l,m} 
= 
 \dfrac{1}{2Q}\,\sum\limits_{q=1}^{Q}\,
 \Big [ &
 \left\vert b^l\left(\vec{s}^l(t_q)\right) - b^m\left(\vec{s}^l(t_q)\right)\right\vert ^2 
\\
+ & 
 \left \vert  b^l\left(\vec{s}^m(t_q)\right) - b^m\left(\vec{s}^m(t_q)\right)\right\vert ^2
\Big].
\end{aligned}
$$
The permutation of control laws $b^l$ and $b^m$ 
with sensors $\vec{s}^l$ and $\vec{s}^m$  
ensures that the distance matrix is symmetric.
More importantly, this ensures that the control laws are compared 
in the relevant sensor space with an averaged probability of both forced attractors.

The resulting distance matrix has the properties that all diagonal elements are equal zero, \textit{i.e.}\ $\mathsfi{D}_{ll}=0,\,\forall l\in\{1,\ldots,N\}$,
and it is symmetric, \textit{i.e.} $\mathsfi{D}_{lm} = \mathsfi{D}_{ml}$.
Applying CMDS to the distance matrix \eqref{Eq:DistanceMatrixControlLaws}, 
each control law $K^l$ is associated with a point $\vec{\gamma}^l$ 
such that the distance between different $\vec{\gamma}^l$ emulates the distance between control laws defined by \eqref{Eq:DistanceMatrixControlLaws}.
More generally, $\vec{\gamma}^l$  are feature vectors which coefficients represent those features 
that contribute most on average to the discrimination of different control laws.

\section{Periodic forcing}
\label{ToC:Periodic forcing}
In this section, periodic forcing (PF) is examined to establish a benchmark for the comparison of control results. The global effects of actuations are discussed in \S~\ref{sec:Actuation effects} from which the optimal actuation setting is obtained. In \S~\ref{sec:nearwake}, near-wake flow characteristics together with the base pressure under the optimal actuation are described. In the following, input and output refer to the experimental plant, \textit{i.e.}, input indicates actuation and output implies sensor.

\graphicspath{{./Figures_pdf/}}
\begin{figure}
	\centering
	\includegraphics[width=0.8\textwidth,keepaspectratio]{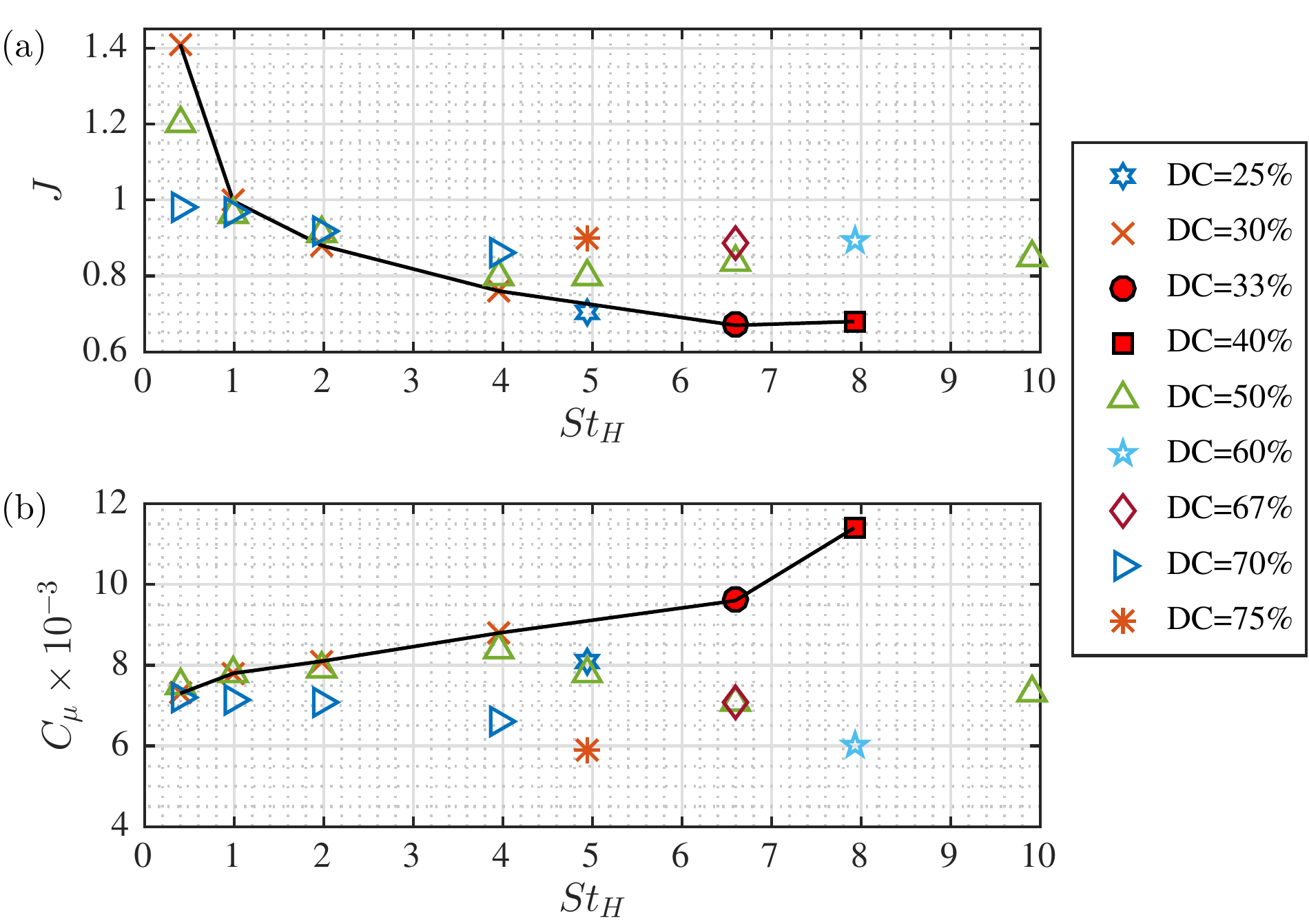}
	\caption{Results of periodic forcing. The variations of $J$ and $C_{\mu}$ versus $St_{H}$ are shown in (\emph{a}) and (\emph{b}), respectively. The initial pressure before actuation is $P_{0}^{i}=\SI{4}{\bar}$. The line connects the configurations where DC$\leqslant$40\%.}
	\label{fig:OL_DC_J_Cu}
\end{figure}

\graphicspath{{./Figures_pdf/}}
\begin{figure}
	\centering
	\includegraphics[width=0.42\textwidth,keepaspectratio]{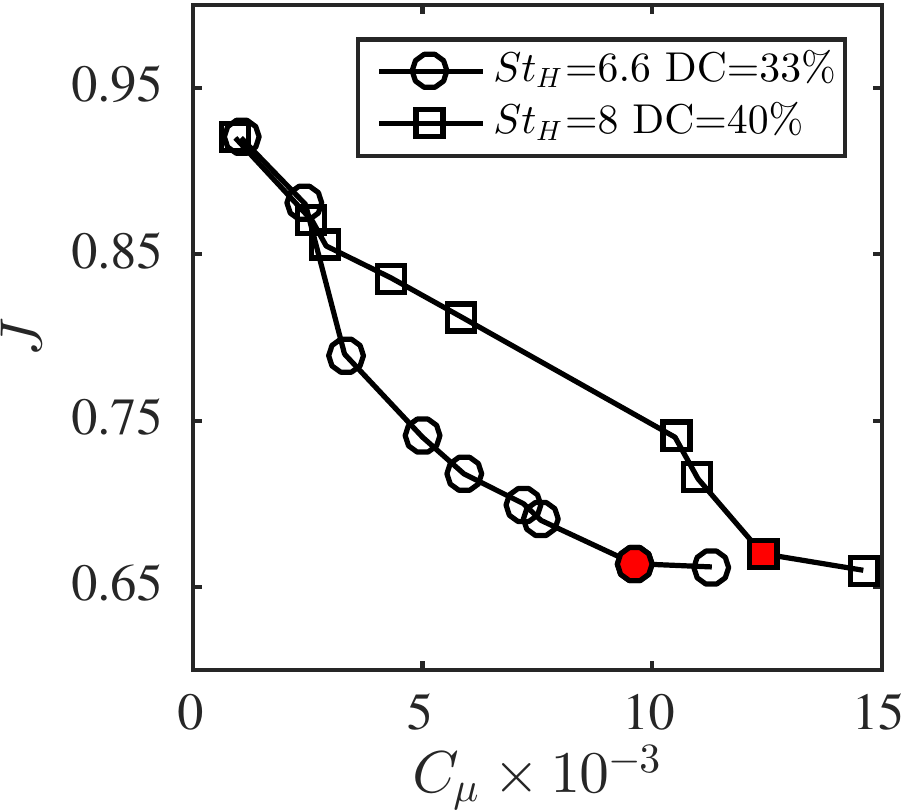}
	\caption{Results of periodic forcing. Variations of $J$ versus $C_{\mu}$ at $St_{H}=6.6$ with $DC=33\%$ and $St_{H}=8$ with $DC=40\%$. The initial pressure before actuation $P_{0}^{i}$ is varied in the range $\left[1.8,4.5\right]\si{\bar}$. The filled symbols display the points corresponding to figure~\protect{\ref{fig:OL_DC_J_Cu}}.}
	\label{fig:J_Cu_333_400}
\end{figure}
%

\subsection{Actuation effects}
\label{sec:Actuation effects}
Except stated otherwise, the same actuation is maintained along all edges. This simultaneous actuation is referred to as single-input. As mentioned in \S~\ref{ToC:Setup}, the frequencies and duty cycles ($DC$) considered for the periodic forcing are the subset of the harmonics derived from $F_{RT}=\SI{2}{\kilo\hertz}$ \textit{i.e.} the red dots in figure \ref{fig:Fre_DC_2kHz}. The actuation amplitude is estimated by the momentum coefficient $C_{\mu}$ defined in (\ref{eq:Cmu}). 

The actuation effects are quantified with respect to the defined cost function $J$. The results are summarized in figure \ref{fig:OL_DC_J_Cu}. The cost is maximized at $St_{H}=0.4$ with DC=30\% resulting in $J=1.4$, which corresponds to 40\% base pressure decrease and 35\% drag increase. According to \citet{Barros2015phd}, the natural vortex shedding appears at $St_{H}=0.2$ corresponding to the typically observed frequencies in the bluff body wakes \citep{roshko1955wake}. The forcing frequency $St_{H}=0.4$ is intriguingly the second harmonic of the natural shedding mode. This behaviour has also been found by \citet{barros2016resonances} with the same model and upstream velocity. Their findings reveal an effect of subharmonic resonance between the forcing and shedding frequencies. This resonance amplifies the vortex shedding oscillations which lead to large base pressure changing and thus decrease the base pressure. 
 
For $1\leqslant St_{H}\leqslant10$, the base pressure increases ($J<1$) for all the values of $DC$. 
In addition, for all the configurations where $DC\leqslant$40\%, $J$ is decreasing as a function of $St_{H}$, 
and $C_{\mu}$ is an increasing function, except for the point at $St_{H}=5$ with $DC$=25\%.
The optimal base pressure recovery is obtained both at $St_{H}=6.6$ and $DC=33\%$ (marked by red filled dot),  and at $St_{H}=8$ and $DC=40\%$ (marked by red filled square) leading to about 33\% recovery of base pressure associated with 22\% drag reduction. 
We note that $C_{\mu}$ in the latter case is higher than the former one, as evidenced in figure \ref{fig:OL_DC_J_Cu}(\emph{b}). To shed some light on the influence of $C_{\mu}$, figure \ref{fig:J_Cu_333_400} shows the variation of the cost $J$ versus a wide range of $C_{\mu}$ 
when the two optimal forcing conditions are used.  
This analysis is performed by tuning the value of the initial pressure $P_0^i$ in the reservoir. 
When $C_{\mu}$ is relatively low, the performance of the two optimal solutions is similar.
For $3.10^{-3}<C_{\mu}<11.10^{-3}$, the performance of the optimal solution with $St_{H}=6.6$ is better than with $St_{H}=8$, the consumed actuation energy being lower for a similar value of $J$.
The results of figure \ref{fig:J_Cu_333_400} indicate that the convergence is already achieved for $St_{H}=6.6$, whereas a slight variation is still visible for $St_{H}=8$.
If we continue to further increase the initial pressure $P_0^i$, the performance of the two optimal solutions will both degrade due to the actuator limitation.

Based on these results, we consider $St_{H}^\star=6.6$ at $DC^\star=33\%$ as the optimal SIPF (single-input periodic forcing) control.
Hereafter, these parameters, denoted $b^\star$, will be used as reference. 
This optimal frequency corresponds to approximately 33 times the one associated to the oscillatory vortex shedding mode determined experimentally. Despite the limited set of frequencies considered here, the base pressure recovery agrees well with that found in \citet{Barros2016jfm} where a wide and refined range of periodic frequencies is studied.

\subsection{Near-wake flow and base pressure}
\label{sec:nearwake}
In this subsection, we discuss the control effects of the optimal SIPF on the near-wake flow and base pressure. 
All physical quantities are normalized by $U_\infty$ and $H$. 

\begin{figure}
	\centering
	\includegraphics[width=1\textwidth,keepaspectratio]{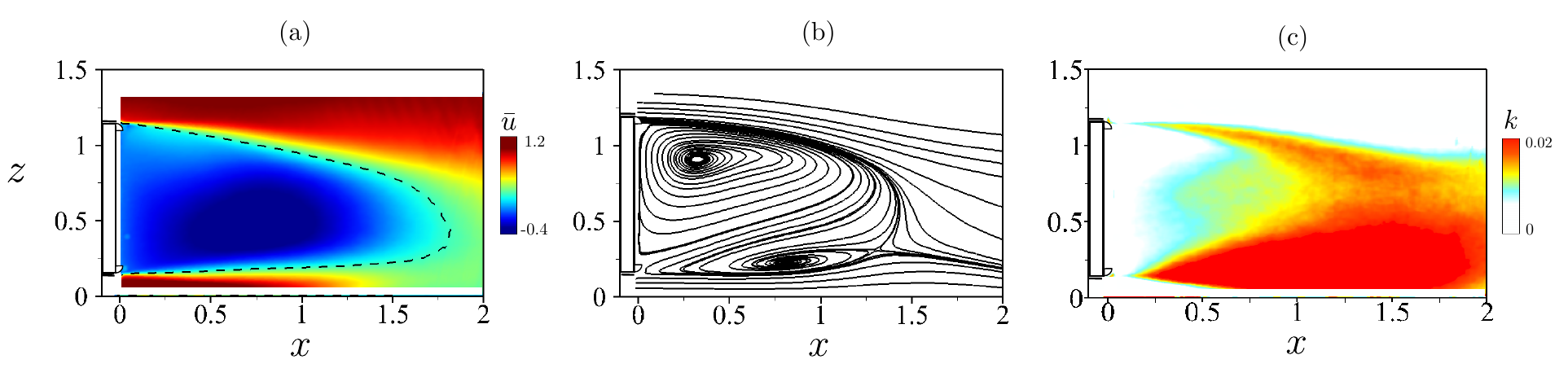}	
	\caption{Time-averaged wake for the unforced flow. From left to right, we present, in the symmetry plane ($y=0$), the contour maps of $\overline{u}$ (\emph{a}), the streamlines (\emph{b}) and the turbulent kinetic energy (\emph{c}).}
	\label{fig:PIV_wake}
\end{figure}

We first give a brief review of the unforced flow to establish a reference for comparison. 
Figure \ref{fig:PIV_wake} shows in the symmetry plane the contour maps of the time-averaged streamwise velocity $\overline{u}$, the streamlines and the turbulent kinetic energy $k=0.5(\overline{u^{{\prime}^2}}+\overline{v^{{\prime}^2}})$ ($u^{\prime}$ and $v^{\prime}$ represent the velocity fluctuations). The dashed line  in figure \ref{fig:PIV_wake}(\emph{a}) indicates the iso-value $\bar{u}=0.25$.  
We remark that the streamlines give only a qualitative 2D picture of the bubble geometry as the flow is fully three-dimensional.
The shear layer emerging from the four leading edges develops and rolls up into large-scale structures. 
This amplification of the shear layer dynamics is crucial to entrain the fluid into the wake region leading to the formation of a recirculation bubble. The negative values of $\overline{u}$ (blue zone) point out clearly the momentum loss in the wake which is closely related to the drag. The dashed line at $\overline{u}=0.25$ provides a reference to the length of the recirculation bubble. 
Streamlines in figure \ref{fig:PIV_wake}(\emph{b}) show that counter-rotating vortices coexist in the wake where the upper structure rotating in the clockwise direction is bigger and closer to the rear surface than the lower one. 
This suggests that the recirculation bubble is dominated by the upper clockwise vortex. 
The vertical wake asymmetry is not surprising as the presence of the ground acts as a perturbation,
leading to flow features that differ from above and under the model. 
The distribution of the turbulent kinetic energy $k$, shown on the rightmost figure, highlights the concentration of $k$ in the shear layer region resulting from the important velocity fluctuations. 
The evolution of the shear layer leads to an increase of $k$ along the streamwise direction. 
Moreover, the fluctuations are more important in the lower shear layer indicating 
that the dynamics is more important near the ground.

\begin{figure}
	\centering
	\includegraphics[width=0.7\textwidth,keepaspectratio]{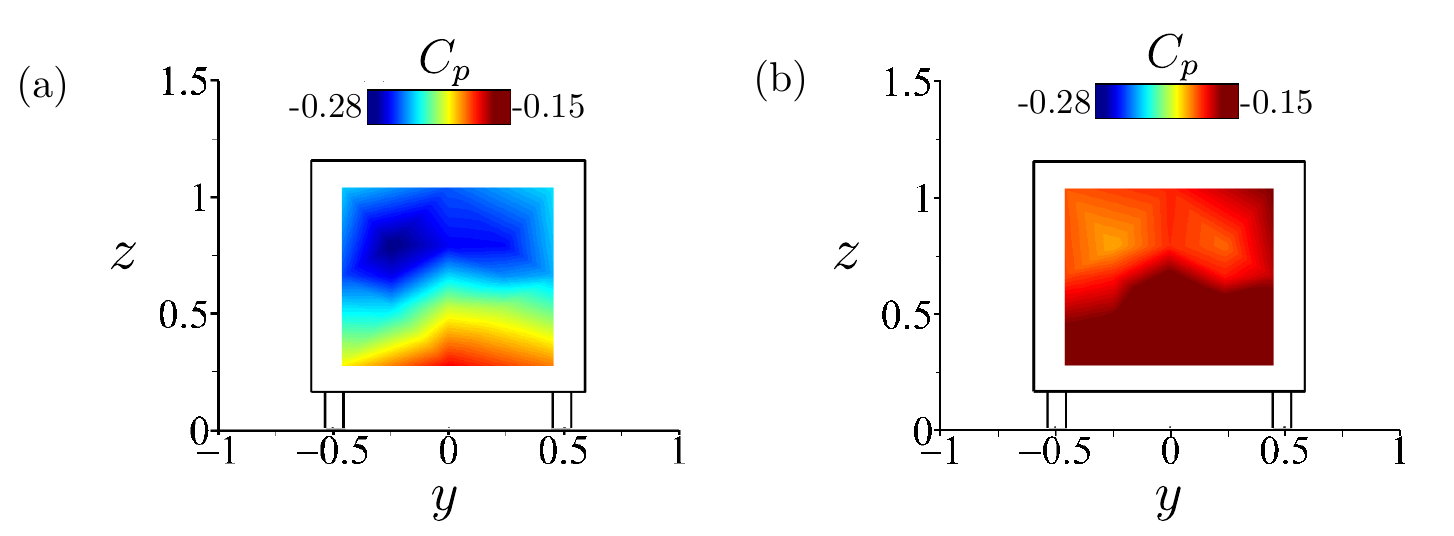}	
	\caption{Time-averaged base pressure for the unforced flow (\emph{a}) and the forced flow with the optimal SIPF (\emph{b}).}
	\label{fig:BasePressureField_Nat_PF}
\end{figure}

The time-averaged base pressure distribution is shown in figure \ref{fig:BasePressureField_Nat_PF}. For the unforced flow, figure \ref{fig:BasePressureField_Nat_PF}(\emph{a}) displays a top-down asymmetry, which is in agreement with the wake topology in figure \ref{fig:PIV_wake}. A low pressure zone is obtained near the upper edge, which is associated with the upper large clockwise vortex. The lateral symmetry (along $y$) can be quantified from the pressure distribution.  Figure \ref{fig:BasePressureField_Nat_PF}(\emph{b}) shows that the base pressure has been globally increased by the forcing. 
The impacts of the high-frequency forcing on the base pressure and wake have been studied thoroughly in \citet{Barros2016jfm} with analyses of wake dynamics. 
The authors explain the mechanism with a two-step process.
First, there is an initial flow deviation close to the separating edges deriving from the \emph{boat-tailing effects} of the combined pulsed jets and Coanda surface. Second, the vortex train generated by the pulsed jets somehow stabilizes the shear-layer growth leading to an overall reduction of kinetic energy and a decrease of entrained flow. The narrowing and more stabilized wake with lower entrainment results in the significant base pressure recovery. 
\graphicspath{{./Figures_pdf/}}

\graphicspath{{./Figures_pdf/}}
\section{Multi-frequency forcing}
\label{ToC:Quasi periodic forcing}

In this section, we extend the search space of open-loop control by looking at multi-frequency forcing constructed by LGPC. The
results of LGPC for this category of actuation are given in \S~\ref{sec:QPF Regression}.
The corresponding optimal control is analysed in detail in \S~\ref{sec:QPF control laws}.

\subsection{LGPC results}
\label{sec:QPF Regression}
In the turbulent wake, the frequency dynamics are broadband suggesting that the periodic forcing space given in \S~\ref{ToC:Periodic forcing} may be not sufficient to search for the optimal control law.
The introduction of multiple frequencies in the actuation should expand the search space of control laws and accommodate this situation. 
LGPC is particularly appropriate for constructing multi-frequency signals and determining the optimal actuation.
First, the actuators are driven in unison. The experimental plant has one single input.
Hereafter, the results of this method are labelled as SIMFF for single-input multi-frequency forcing.

In the LGPC framework, the open-loop control laws can be written as $b(t)=K({\vec{h}(t)})$ where $\vec{h}$ is the input vector of harmonic control laws. We define $\vec{h}=\{h_1,...,h_{9}\}$ where $h_i(t)=\sin(2\pi f_i t)$ represents the harmonic function at the frequency $f_i$.
The values of $f_i$ considered in this study and the corresponding Strouhal number $St_{H_i}=f_i H/U_\infty$ are presented in table \ref{tab:h_i}. The goal is to find an optimal function $K^\text{h}$, where the superscript $\text{h}$ indicates harmonic, such that $b^\text{h}(t)=K^\text{h}({\vec{h}(t)})$ minimizes the cost function $J$.

\begin{table}
	\centering
	\begin{tabular}{l*{8}{c}r}
		Controller input  & $h_1$& $h_2$& $h_3$& $h_4$& $h_5$& $h_6$& $h_7$& $h_8$& $h_9$  \\
		$f_i$ (\si{\hertz})  & 10& 20& 50& 100& 200& 250& 333& 400& 500  \\
		$St_{H_i}$  & 0.2& 0.4& 1& 2& 4& 5& 6.6& 8& 10  \\
	\end{tabular}
	\caption{Description of the harmonic functions $h_i(t)=\sin(2\pi f_i t)$ used as inputs of LGPC for multi-frequency forcing.}
	\label{tab:h_i}
\end{table}

\graphicspath{{./Figures_pdf/}}
\begin{figure}
    \centering
	\subfigure{
		\includegraphics[width=0.72\textwidth,keepaspectratio]{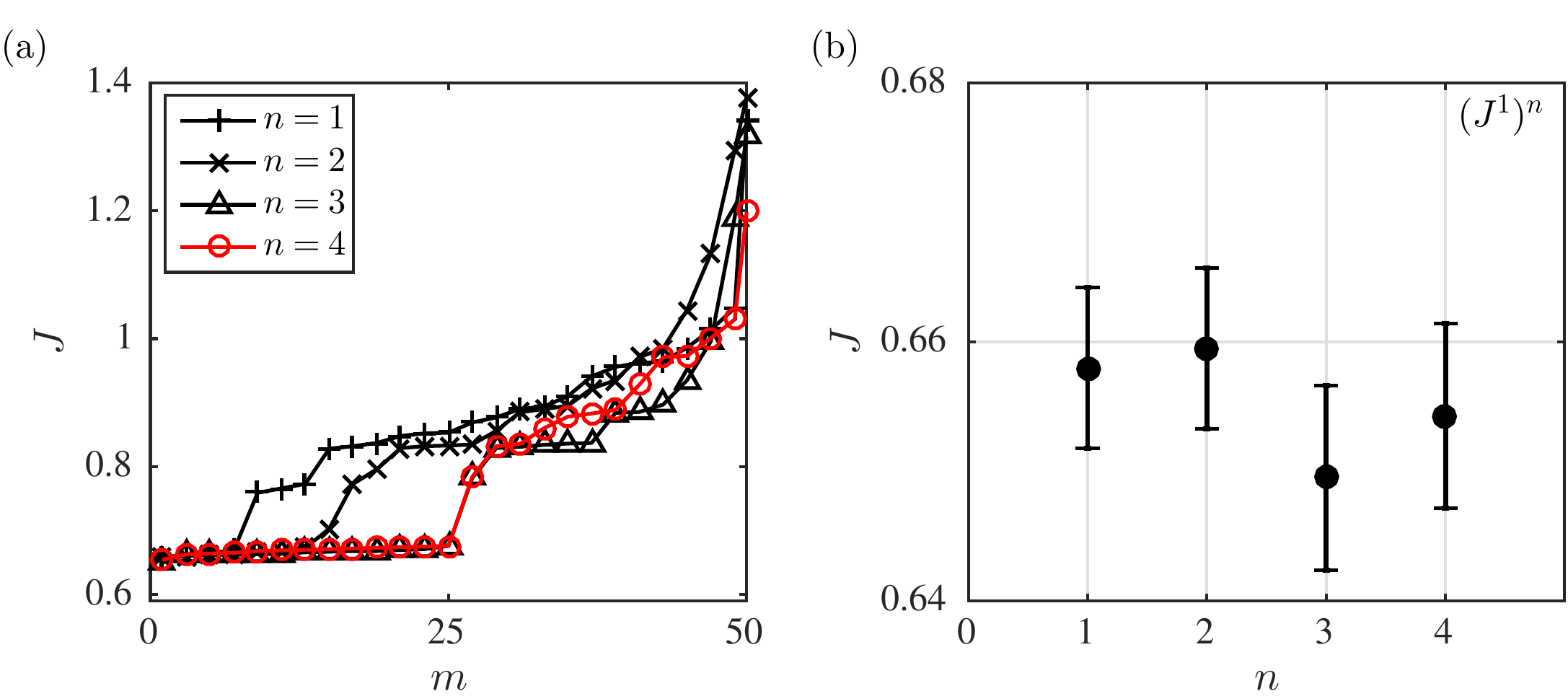}}
	\vfill
	\centering
	\subfigure{
		\includegraphics[width=1\textwidth,keepaspectratio]{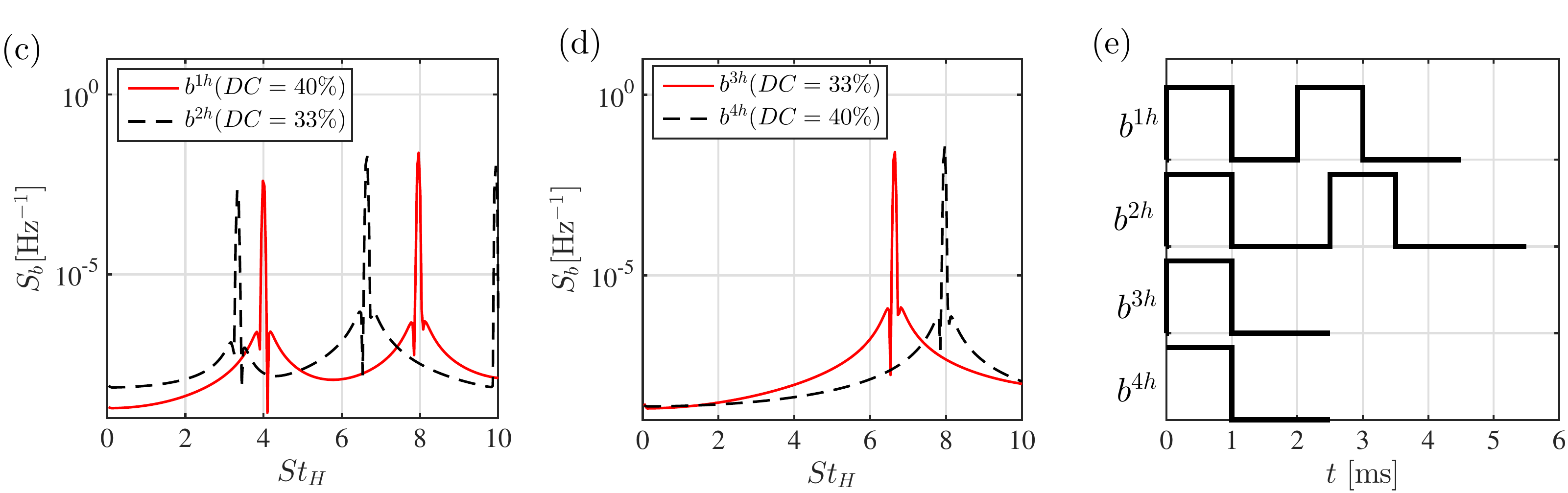}}
	\caption{%
	Results of LGPC for single-input multi-frequency forcing. 
	(\emph{a}) Evolution of the cost function $J$ versus the individuals $m$ for four generations $n=1,\ldots,4$. 
	(\emph{b}) Cost of the optimal individual in the first four generations $(J^1)^n,\,n=1,\ldots,4$. 
	Power spectral density of the control law ($S_{b}$) for $b^{1\text{h}}$, $b^{2\text{h}}$ (\emph{c}) and $b^{3\text{h}}$, $b^{4\text{h}}$ (\emph{d}). (\emph{e}) One period of the actuation $b^{m\text{h}}$ ($m=1,\ldots,4$).}
	\label{fig:QPF_LGP}
\end{figure}

The results are summarized in figure \ref{fig:QPF_LGP}. 
The evolution of $J$ versus the number of the individual $m=\{1,...,M\}$, is displayed in figure \ref{fig:QPF_LGP}(\emph{a}) for the four generations $n={1,...,4}$. 
Most of the control laws are effective ($J<1$) from the first generation. 
This is in agreement with the results of SIPF where only a narrow range of frequencies results in $J>1$. 
LGPC is stopped at the fourth generation because half of the individuals have similar $J$ values near the optimal one. 
When the number of generations increases, we observe a global trend to obtain lower values of $J$, but the evolution of the top-performing individuals is slight. 
To illuminate this behavior, the cost $(J^{1})^n$ of the optimal individual in each generation $n$ is shown in figure \ref{fig:QPF_LGP}(\emph{b}).
The dots correspond to the averaged $J$ values and the error bars show the standard deviation of repeating evaluations of the optimal control law.
Due to the experimental uncertainties, 
the dots are not strictly monotone.
For the two first generations ($n=1,2$), the optimal individual exhibits the same frequency ($St^\star_{H}=6.6$) and duty cycle ($DC^\star$=33\%) as the optimal SIPF solution $b^\star$.
In the third generation, a new individual evolves leading to a gain of 1\% in reduction of $J$ (about 3\% relative benefit) compared to the previous generations.
At the fourth generation, this individual does not evolve favourably in average.   
Now, we focus on the converged generation ($n=4$).
Due to the binary ON/OFF command, and the necessity to apply an Heaviside function to the control law (see \S~\ref{sec:LGP}), it is 
possible to have different control laws which give the same actuation $b$.
For instance, the first five individuals may have only three kinds of actuations.
In the following, we name $b^{1\text{h}}, b^{2\text{h}}, b^{3\text{h}}$ and $b^{4\text{h}}$, the first four distinct actuations in the top-ranking individuals of generation $n=4$.
The actuation power spectral densities $S_b$ for $b^{m\text{h}}$ ($m=1,\ldots,4$) are displayed in figure \ref{fig:QPF_LGP}(\emph{c}) and (\emph{d}) in the range $St_{H}\in[0,10]$. 
One period of the actuation $b^{m\text{h}}$ is presented in figure \ref{fig:QPF_LGP}(\emph{e}).
The solutions $b^{3\text{h}}$ and $b^{4\text{h}}$ contain a single-frequency corresponding to the two optimal solutions found in \S~\ref{ToC:Periodic forcing}, \textit{i.e.} $St^\star_{H}=6.6$ with $DC^\star$=33\% and $St_{H}=8$ with $DC$=40\%, respectively. 
$b^{1\text{h}}$ and $b^{2\text{h}}$ exhibit a multi-frequency dynamics with the dominant frequency of $b^{4\text{h}}$ ($St_{H}=8$) and $b^{3\text{h}}$ ($St^\star_{H}=6.6$), respectively, and their subharmonic ($St_{H}=4$ and $St_{H}=3.3$).  

\begin{table}
	\centering
	\begin{tabular}{l*{4}{c}}
		Control law  & $J$  & $C_\mu(\times10^{-3})$ & $A_e$ & $P_s$ \\~\\
		
		$b^{1\text{h}}=\mathcal{H}\left({h_{5}}/{h_{8}}-0.622\right)$ & 0.654 & 9.834 & 2.958 & 0.156\\
		
		$b^{2\text{h}}=\mathcal{H}\left((h_{9}-h_{7}-0.2\right)$ & 0.661 & 10.927 & 2.317 & 0.131 \\
		
		$b^{3\text{h}}=\mathcal{H}\left((-0.479h_{7}-0.2\right)$ & 0.664  & 9.609 & 2.841 & 0.146 \\
		
		$b^{4\text{h}}=\mathcal{H}\left((\tanh(-h_8/0.376)-0.2\right)$  & 0.668  & 11.379 & 2.122 & 0.110\\
	\end{tabular}
	\caption{LGPC optimal control laws for the single-input multi-frequency forcing. We give the performance of the four top performing individuals in the last generation ($n=4$). $\mathcal{H}$ represents the Heaviside function. By definition, $\mathcal{H}(x)=0\text{ if } x\leqslant0$ and $\mathcal{H}(x)=1$ otherwise.}
	\label{tab:QPF_LGP_table}
\end{table}
%

To discuss the energetic efficiency of the control, we define an actuation efficiency coefficient $A_e$ and a relative power savings coefficient $P_s$ as follows:
\begin{equation}
A_e=\dfrac{|\Delta{C_{D}}| S U_\infty^3}{S_\text{Jet}\overline{V_\text{Jet}^3}}~\text{ and }~
P_s=\dfrac{\frac{1}{2}|\Delta{C_{D}}|SU_\infty^3-\frac{1}{2}S_\text{Jet}\overline{V_\text{Jet}^3}}{\frac{1}{2}C_D^\text{u}SU_\infty^3}, 
\label{eq:equa01}
\end{equation}
with $\Delta{C_{D}}=C_{D}^\text{u}-C_{D}^\text{f}$ where the superscripts u and f indicate the unforced and forced flows, respectively.
The actuation efficiency $A_e$ represents the ratio between the mechanical power gained by the drag reduction and the mechanical power consumed by the pulsed jets.
The relative power saving $P_s$ represents the net power saving related to the control normalized by the power consumed by the aerodynamic drag in the unforced flow.
The expressions of the control laws $b^{m\text{h}},\,m=(1,...,4)$ are reported in table \ref{tab:QPF_LGP_table} with the corresponding values of the cost $J$, the actuation amplitude $C_{\mu}$, the actuation efficiency $A_e$ and the power saving $P_s$.
All the actuation efficiencies $A_e$ are greater than 1 and all the relative power saving coefficients $P_s$ are greater than 0 indicating that the net energy balance is positive.
The optimal actuation $b^{1\text{h}}$ results in about 34.6\% of base pressure recovery. 
The returned gain of the invested actuation power is approximately three. 
The power consumed by the aerodynamic drag has been saved by 15.6\%.

\graphicspath{{./Figures_pdf/}}
\begin{figure}
	\centering
	\includegraphics[width=0.8\textwidth,keepaspectratio]{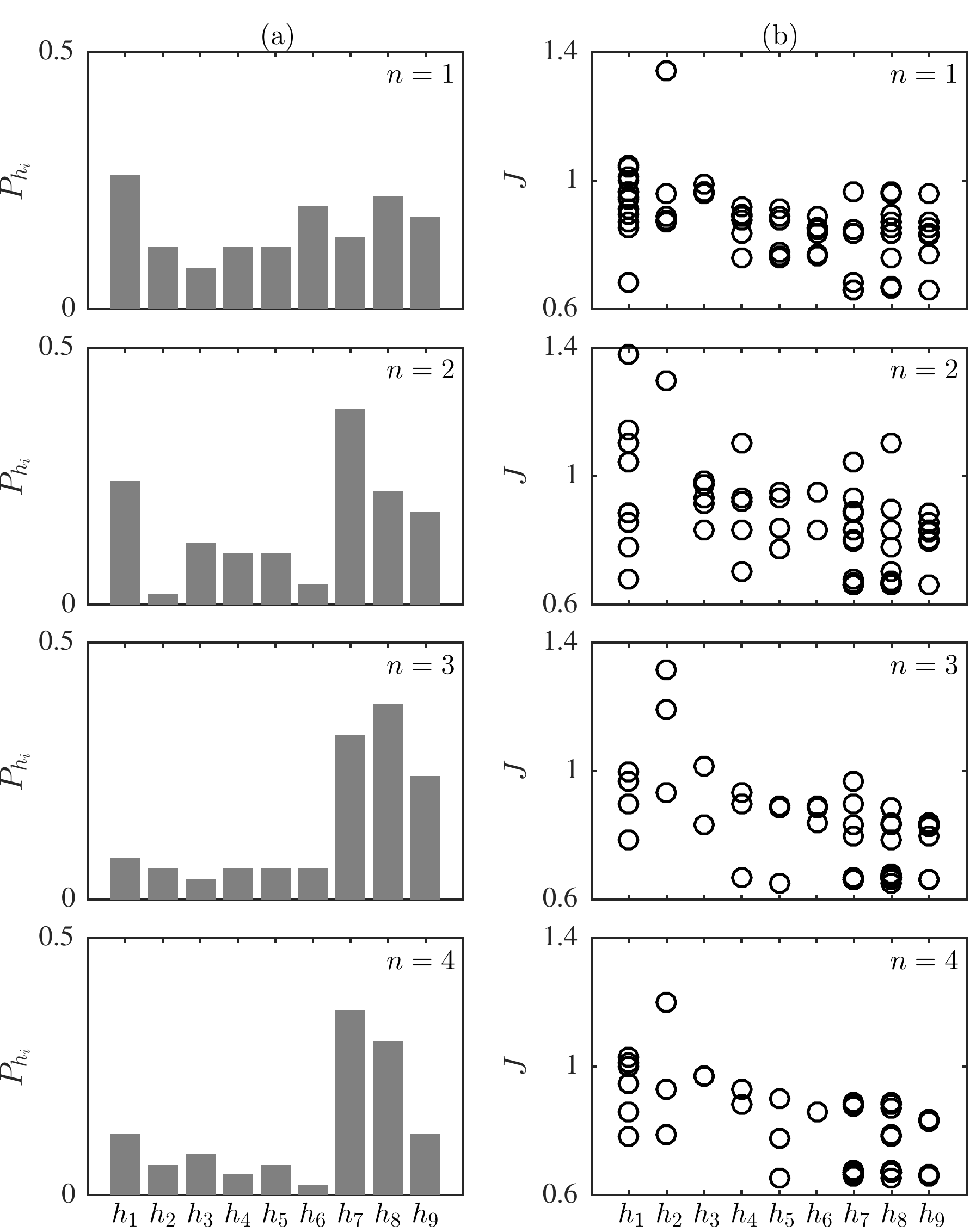}	
	\caption{%
		Convergence of LGPC for single-input multi-frequency forcing.
		For each generation $n=1,...,4$, we represent:
		(\emph{a}) the percentage $P_{h_i}$ of having $h_i$ in the expression of the individuals,
		(\emph{b}) the spectrum of $J$-value of individuals which include $h_i$ in their expression.}
	\label{fig:PIMO_Distribution_s_J}
\end{figure}

In the following, we study the convergence of LGPC towards the optimal control law by analysing how the different harmonic functions $h_i$ are selected over the generations. 
The percentage $P_{h_i}$ of having $h_i$ involved in the individuals is displayed in figure \ref{fig:PIMO_Distribution_s_J}(\emph{a}) for all the generations.
In the first generation, all the harmonic functions are loosely equivalent to be chosen.
Note that if the population was larger, we would have a uniform distribution of $h_i$.
At the second generation, the percentage of presence of $h_7$ raises abruptly for becoming largely dominant.
At the third generation, $P_{h_8}$ grows and now becomes dominant.
At convergence, $h_7$ and $h_8$ are the two harmonic functions most commonly found in all the individuals.
This result, perfectly in agreement with the optimal frequencies found in \S~\ref{ToC:Periodic forcing}, demonstrates the ability of LGPC to select automatically the optimal harmonic forcing parameters.
The spectrum of $J$-values of individuals which include $h_i$ in the expression of the individuals is showed for all the generations in figure \ref{fig:PIMO_Distribution_s_J}(\emph{b}).
More precisely, we plot for each individual the cost $J$ (in ordinate) against the harmonic functions $h_i$ (in abscissa) occurring in it.
Over the generations, the data points move progressively from a relatively sparse distribution to a concentrated distribution in the bottom right region, proving that the best individuals are obtained for high frequencies.  

In addition, multiple-input multi-frequency forcing (MIMFF) has also been performed by driving the top, down, left and right actuators independently. The search space is much more larger than that of single-input control. 
The optimal single-input control law (SIMFF) was inserted in the first generation to accelerate convergence. 
LGPC with multiple inputs did not improve the performance for the best single-input law.

\subsection{Analysis of the optimal control law}
\label{sec:QPF control laws}
\begin{figure}
	\centering
	\includegraphics[width=1\textwidth,keepaspectratio]{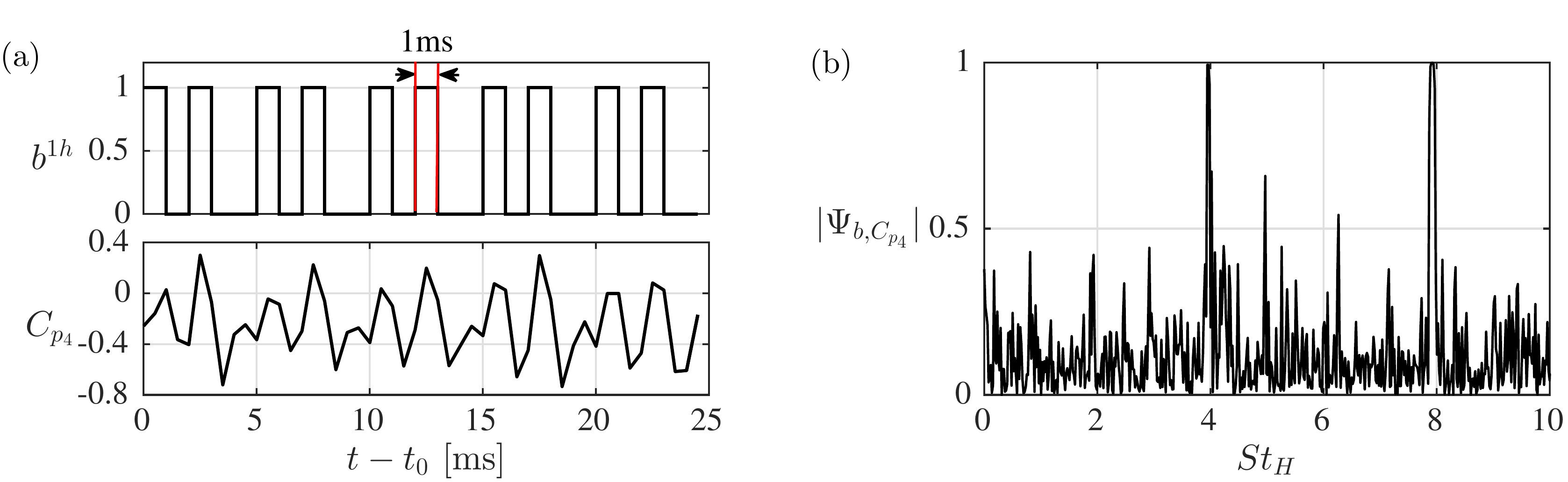}	
	\caption{Impact of the optimal SIMFF $b^{1\text{h}}$ on the pressure coefficients. (\emph{a}) Instantaneous response of $C_{p_4}$ to the actuation $b^{1\text{h}}$. The time is shifted by a randomly chosen value $t_0$. (\emph{b}) Amplitude of the spectral coherence between the actuation $b^{1\text{h}}$ and pressure coefficient $C_{p_4}$.}
	\label{fig:PIMO_Corrf}
\end{figure}

In this subsection, we focus on how the optimal single-input multi-frequency forcing (SIMFF) $b^{1\text{h}}=\mathcal{H}\left({h_{5}}/{h_{8}}-0.622\right)$ influences the base pressure. For that, we first investigate the instantaneous impact of actuation on the base pressure. Figure \ref{fig:PIMO_Corrf} (\emph{a}) represents the time evolution of the bottom-middle pressure coefficient $C_{p_4}$ under several periods of the actuation $b^{1\text{h}}$. $C_{p_4}$ is corrected in amplitude and phase based on the approach described in \S~\ref{sec:Pressure_measurements}. In addition, $b^{1\text{h}}$ is shifted $\SI{1}{\milli\second}$ downward in time to take into account the actuator delay (see \S~\ref{sec:actuator}). 
The corrected signals are used here because we are interested in the pressure response at the base surface to the actuation.
We notice in figure \ref{fig:PIMO_Corrf} (\emph{a}) that the apparent frequency in pressure fluctuation is tightly related to that of actuation. 
This correlation can be further inferred from the spectral coherence $\Psi_{b,C_{p_i}}$ which is defined at each frequency $f$ as follows:
\begin{equation}
\Psi_{b,C_{p_i}}(f)=\frac{G_{b,C_{p_i}}(f)}{\sqrt{G_{b}(f)\,G_{C_{p_i}}(f)}},\quad i={1,...,16}
\label{eq:spectral coherence}
\end{equation}
where $G_{b,C_{p_i}}$ is the cross-spectral density between the actuation $b^{1\text{h}}$ and $i$th pressure coefficient $C_{p_i}$, and $G_{b}$ and $G_{C_{p_i}}$ are the auto-spectral density of $b^{1\text{h}}$ and $C_{p_i}$, respectively. Figure \ref{fig:PIMO_Corrf}(\emph{b}) displays the amplitude of the spectral coherence $\Psi_{b,C_{p_i}}$. We observe a level of coherence of about 100\% at $St_{H}=4$ and $St_{H}=8$ which are indeed the forcing frequencies shown in figure \ref{fig:QPF_LGP}(\emph{c}). 
These high values of coherence at the forcing frequencies have been equally observed for the other pressure signals implying that all sensors over the base are correlated to the actuation regardless of their locations. 
From these observations a question arises: do the sensors respond to the actuation at the same time? To address this question, the coherence $\Psi_{C_{p_i},C_{p_j}}$ between the pressure signals $C_{p_i}$ and $C_{p_j}$ is studied. 
From $\phi_{i,j}$ the phase of $\Psi_{C_{p_i},C_{p_j}}$, we have determined the time shift at frequency $f$
between the pressure signals $C_{p_i}$ and $C_{p_j}$ as $\phi_{i,j}/(2\pi f)$.
This value is of the order of \SI{0.2}{\milli\second} which may be related to the distance between pressure taps and to the slight length difference of tube mounting of two sensors. 
We can then conclude that all the pressure signals respond to the actuation at the same time.

\graphicspath{{./Figures_pdf/}}

The results above have important implications for the understanding of actuation effects. As described at the end of \S~\ref{sec:nearwake}, the combination of pulsed jets and Coanda surface creates a \emph{boat-tailing effect} resulting in an inward shear layer deviation close to the separating edges, and thus yields a time-averaged base pressure increase. 
Here we want to elucidate the existence of an instantaneous boat-tailing effect by analysing the temporary response of the pressure to the actuation. 
The underlying dynamics can be derived from the time history of spatially averaged pressure $\langle C_p\rangle$ under the unsteady forcing.
\graphicspath{{./Figures_pdf/}}
\begin{figure}
 	\centering
 	\includegraphics[width=0.7\textwidth,keepaspectratio]{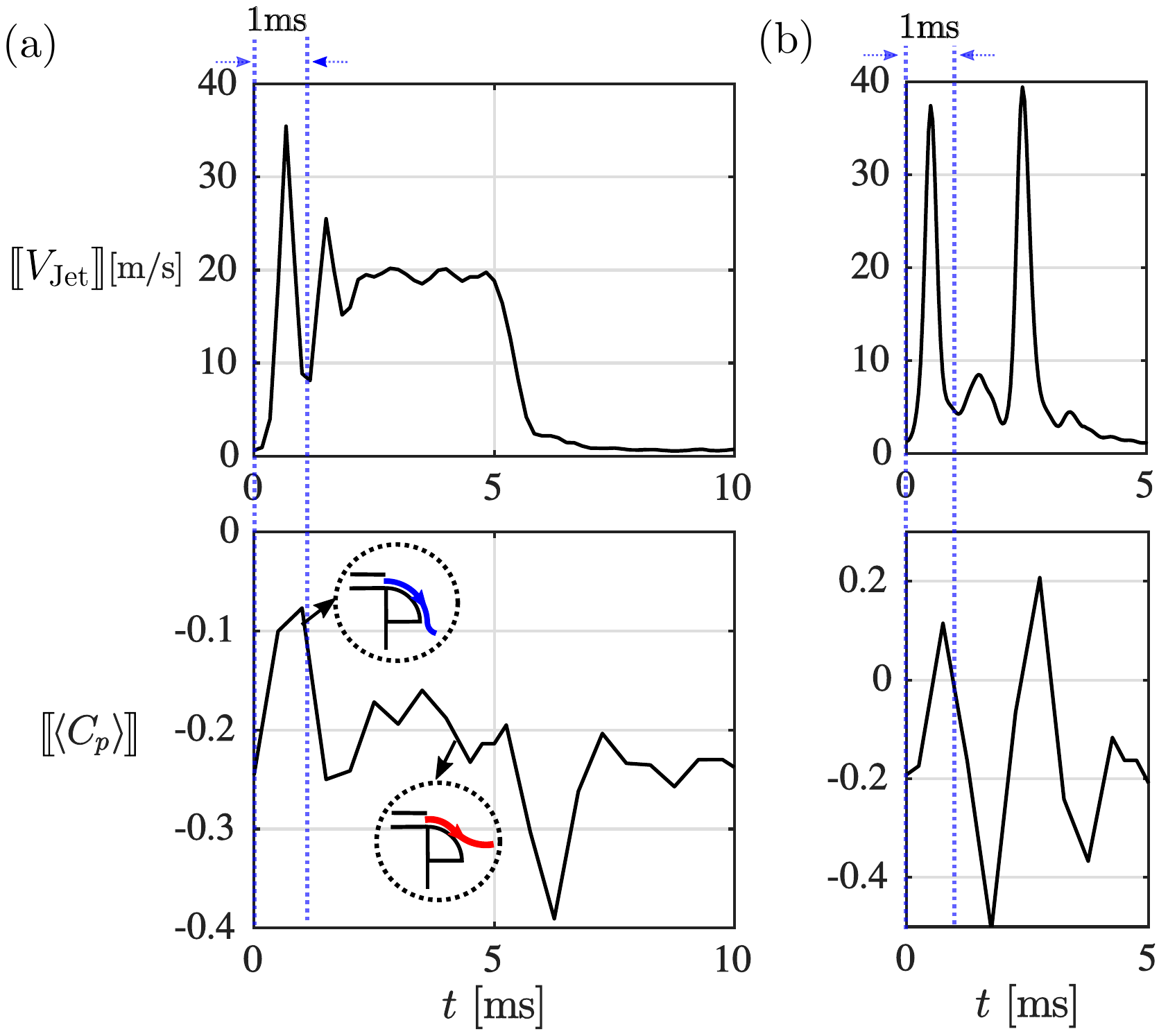}
 	\caption{Phase-averaged jet velocity $\llbracket V_\text{Jet}\rrbracket$ and spatially phase-averaged pressure coefficient $\llbracket\langle C_p\rangle\rrbracket$. (\emph{a}) Forced flow at $St_{H}=2$ and $DC=50\%$. (\emph{b}) Optimal SIMFF $b^{1\text{h}}$. The phase average is performed with respect to the lower frequency \textit{i.e.} $St_{H}=4$. The dashed line indicates the time duration for the unsteady overshoot. The inserted figures indicate the different interaction of the jet flow with the Coanda surface in the unsteady and quasi-steady state, respectively.}
 	\label{fig:Unsteady_effect_Vj_Cp}
\end{figure} 
 Figure \ref{fig:Unsteady_effect_Vj_Cp} shows the phase-averaged jet velocity $\llbracket V_\text{Jet}\rrbracket$ and spatially phase-averaged coefficient $\llbracket\langle C_p\rangle\rrbracket$ under a moderate frequency $St_{H}=2$ (\emph{a}) and the optimal SIMFF control $b^{1\text{h}}$ (\emph{b}). A moderate value of frequency was chosen in (\emph{a}) to gain insights on the jet propagation over the surface due to its relatively long pulse duration.  
An overshoot of $\llbracket V_\text{Jet}\rrbracket$ is observed at the very beginning of the blowing in figure \ref{fig:Unsteady_effect_Vj_Cp}(\emph{a}), which is related to a sudden pressure relief in the reservoir when the valve opens. This overshoot lasts about \SI{1}{\milli\second} and then the jet stabilizes and develops to a quasi-steady blowing. A similar duration of the overshoot has been observed in lower frequencies indicating that this is probably a characteristic of the actuators. 
The pressure signal shows correspondingly a sudden and strong increase just at the same time of the overshoot. Following the stabilization of the blowing, the pressure also stabilizes and fluctuates around a particular value. We conjecture that the fluctuations at the different states, unsteady overshoot and quasi-steady blowing, are related to the movement of the separation point over the Coanda surface. We then propose a conceptual scenario attempting to explain the different mechanisms in the unsteady overshoot and quasi-steady state. 
During the unsteady overshoot, the jet travels over the rounded surface carrying a strong velocity inside the forefront of the jet while facing a relatively low-velocity flow on its outside. As a first-order approximation, this process is too short to give the opportunity to the viscosity to affect the flow. Therefore, the instantaneous velocity acceleration is almost totally used to compensate the reversed pressure force over the rounded surface.   
We denote by $t_\text{Prop}$ the propagation time of the jet from the slit exit to the end of the Coanda surface. This propagation time can be estimated as $t_\text{Prop}=\ell/\overline{V_\text{Jet}}$ where $\ell=\pi r/2$ is the arc length of the surface and $\overline{V_\text{Jet}}$ is the time-averaged jet velocity. 
If we approximate $\overline{V_\text{Jet}}$ by the oncoming velocity $U_{\infty}$, we get $t_\text{Prop}=\SI{0.94}{\milli\second}$. This value is surprisingly close to the duration of the unsteady overshoot. 
This means that the jet flow can completely attach on the surface within the unsteady state under the condition of $\overline{V_\text{Jet}}>U_{\infty}$.
By intuition, the flow may be highly deviated as illustrated in the inserted figure for the overshoot state.
Once entering into the steady state, the pressure fluctuation decreases. The reason is two-fold:
first, the velocity jet has significantly decreased compared with the overshoot resulting in a lower velocity acceleration; second, the jet momentum dissipates with the increasing time due to the viscous effects. The flow resistance is lower to the reversed pressure gradient on the surface. Both of them lead to an earlier flow separation and a less deviated flow, as shown in the  inserted figure for the quasi-steady state. When the blowing is stopped, there is a significant decrease of pressure which remains unclear. Everything happens as if the jet closure somehow induces a strong detachment of the flow.

Given the important role played by the unsteady effect, one would expect that the actuation should take advantage of this unsteady overshoot to gain benefits in the base pressure. 
Figure \ref{fig:Unsteady_effect_Vj_Cp}(\emph{b}) shows the phase-averaged jet velocity $\llbracket V_\text{Jet}\rrbracket$ measured for the actuation $b^{1\text{h}}$. This jet velocity exhibits two overshoots in one period. The base pressure is consequently excited to a high value and leads to an ultimate time-averaged base pressure recovery.
This explains why the high-frequency forcing yields a better performance. 
We define $t_\text{Pulse}$ the pulse duration of one pulsed jet and $t_\text{Int}$ the intermittent time between two successive pulsed jets. It is expected that $t_\text{Pulse}$ could be as small as possible to eliminate the quasi-steady blowing. In addition, $\llbracket V_\text{Jet}\rrbracket$ should be strong enough to drive the jet to the end of the Coanda surface in $t_\text{Pulse}$. 
Considering the actuator response time and the characteristic time for the overshoot, the smallest value for $t_\text{Pulse}$ is determined to be $\SI{1}{\milli\second}$. 
Surprisingly, the top-ranking individuals in figure \ref{fig:QPF_LGP}(e) are all in good agreement with our hypothesis. They have all $t_\text{Pulse}=\SI{1}{\milli\second}$ but $t_\text{Int}$ is different. The optimal control $b^{1\text{h}}$ is the only actuation including $t_\text{Int}=\SI{1}{\milli\second}$, as shown in figure \ref{fig:PIMO_Corrf}(\emph{a}). We may conclude that $b^{1\text{h}}$ meets best the requirement for $t_\text{Pulse}$, $t_\text{Int}$ and $\llbracket V_\text{Jet}\rrbracket$ and therefore is chosen as the optimal controller. 


\section{Feedback control}
\label{ToC:Feedback control}
In this section, we explore the opportunities of sensor-based closed-loop control. 
First (\S~\ref{sec:aSIMO}), 
a single-input multiple-output (SIMO) control system is studied. 
Filtered time-history feedback is considered in \S~\ref{sec:MI_LGPC} for both single-input and multiple-input control. 
In \S~\ref{sec:other_cases}, 
we combine sensor feedback and the optimal periodic forcing in 
a general non-autonomous closed-loop control.

\subsection{Single-input multiple-output control}
\label{sec:aSIMO}
In this subsection, all actuators are operated simultaneously by a single actuation command.
The results of LGPC are given in \S~\ref{sec:aSIMO_Regression}. 
The resulting control laws are visualized and interpreted in \S~\ref{sec:aSIMO_visualization}.
Section~\ref{sec:aSIMO_control_laws} presents a physical analysis of the optimal control law.

\subsubsection{LGPC results}
\label{sec:aSIMO_Regression}
The closed-loop control law is expressed as $b=K(\vec{s})$,
where $\vec{s}$ consists of the pressure sensors distributed over the rear surface. 
For the feedback, it was found that 
the first 12 sensors in figure \ref{fig:sketch_setup}(\emph{c}) are sufficient for the performance of the controller, 
the sensors 13--16 providing redundant information.  
The cost $J$ is evaluated based on all 16 sensors following \eqref{eq:J}.
This control is referred as single-input multiple-output (SIMO) 
as we have one actuation command and 12 sensor signals. 
From the sensors, only the fluctuation part is fed back to mitigate the effect of slow drifts. 
The fluctuation of $i${th} sensor $s^\prime_{i}$ is defined as:
\begin{equation}
s^\prime_i(t)=s_i(t)-\overline{s_i}(t)
\end{equation}
where
\begin{equation}
\overline{s_i}(t) =\frac{1}{\tau_\text{\tiny av}}\int_{t-\tau_\text{\tiny av}}^{t}s_i(t)\,\text{d}t
\end{equation}
is the moving average of the signal over a period $\tau_\text{\tiny av}=\SI{0.1}{\second}$. 
Summarizing,  the control law has the form 
$$b^{s}=K^{s}(\vec{s}^\prime) \quad \hbox{with}\quad   \vec{s}^\prime=\{s^\prime_1,...,s^\prime_{12}\}$$
where the superscript $s$ indicates sensor-based control law. 
The number of sensors being greater than in \S~\ref{ToC:Quasi periodic forcing}, 
we increased the number of instructions in the individuals to a range varying from 20 to 50.
The other LGPC parameters remain the same.

\graphicspath{{./Figures_pdf/}}
\begin{figure}
	\centering
		\includegraphics[width=1\textwidth,keepaspectratio]{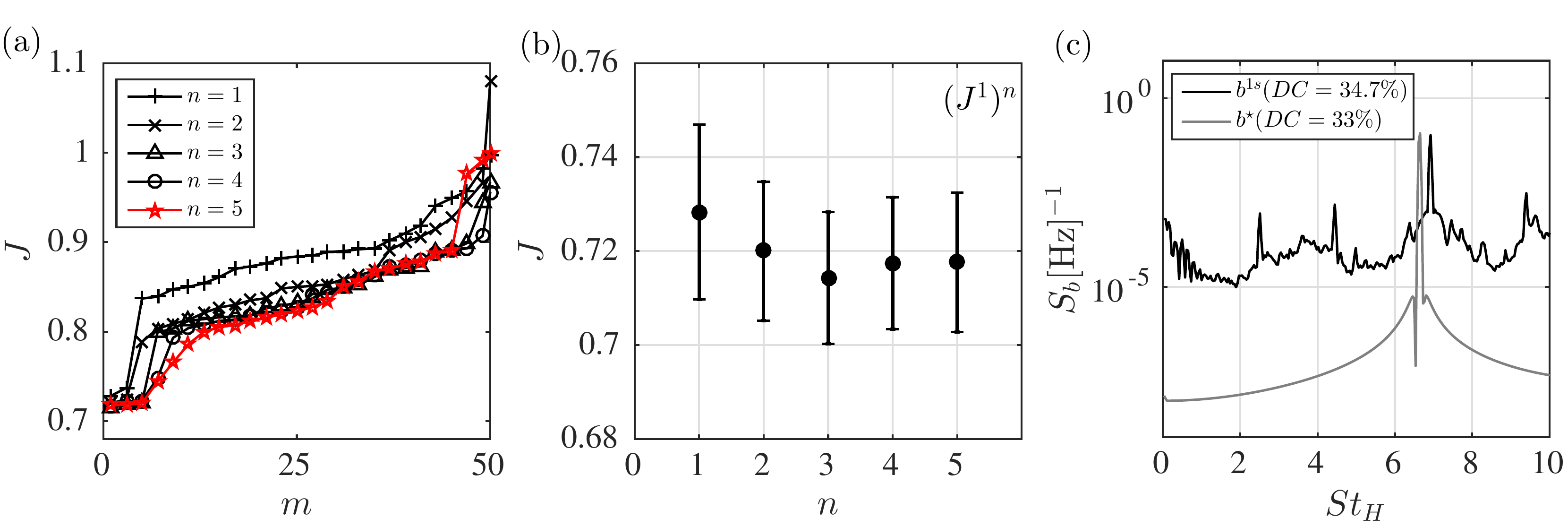}
	\caption{Results of LGPC for sensor-based single-input multiple-output (SIMO) control. (\emph{a}) Evolution of the cost function $J$ versus the individuals $m$ for five generations $n=1,\ldots,5$. (\emph{b}) Cost of the optimal individual $(J^1)^n$ in each generation $n$. (\emph{c}) Power spectral density $S_{b}$ for the optimal SIMO $b^{1s}$ and the optimal SIPF $b^\star$.}
	\label{fig:aSIMO_LGP}
\end{figure}

The results of the LGPC experiment are presented  in figure \ref{fig:aSIMO_LGP} 
in a similar way as figure \ref{fig:QPF_LGP}. 
We stop LGPC after $n=5$ generations 
because the cost $J$ does not evolve anymore. 
Figure~\ref{fig:aSIMO_LGP}(\emph{a}) shows the evolution of $J$ 
versus the index of the individual $m$. 
Almost all the individuals improve their values of cost function compared to those of the first generation.   
We focus on the evolution of the optimal individual in each generation in figure \ref{fig:aSIMO_LGP}(\emph{b}). 
The optimal individual yielding $J\approx0.72$ is found from the generation $n=2$ and is further confirmed as the optimal one until $n=5$. 
The error bar is determined from re-evaluations of mathematically equivalent control laws in all the generations.
Most actuation commands extract high-frequency components from the sensor signals. 
The reason may be two-fold:
first, the individuals with high-frequency forcing are more evolved by LGPC 
since they can lead to better $J$-values;
second, some low frequencies are filtered out by the moving average of the signal.
The spectrum of the optimal individual in the final generation $n=5$, named as $b^{1s}$, 
is shown in figure~\ref{fig:aSIMO_LGP}(\emph{c}). 
The spectrum of the optimal SIPF $b^{\star}$ is also included for comparison. 
$b^{1s}$ evidences a dominant frequency at $St_{H}=6.9$ with duty cycle of $DC=34.7\%$.
Both parameters are quite close to those of $b^{\star}$. 
However, $b^{1s}$ has a richer spectrum than $b^{\star}$.

\begin{table}
	\centering	
	\begin{tabular}{l*{6}{c}}
		&$St_{H}$ & DC & $J$  & $C_{\mu}(\times10^{-3})$ & $A_e$ & $P_s$\\	
		$b^{1s}$ & 6.9 & 34.7 & 0.718 & 10.147 & 2.428 &  0.096\\
		$b^{\star}$ & 6.6 & 33 & 0.664 & 9.609 & 2.841 & 0.146 \\
	\end{tabular}
	\caption{Performance of the optimal SIMO control $b^{1s}$ compared with the optimal SIPF $b^{\star}$.}
	\label{tab:aSIMO_table}
\end{table}

Table \ref{tab:aSIMO_table} compares the main characteristics of $b^{1s}$ and $b^{\star}$. 
Closed-loop control has similar actuation features (dominant frequency and duty cycle) as $b^{\star}$.
Yet, the performance of $b^{1s}$ is slightly worse. 
The presence of low-frequency components in $b^{1s}$ may degrade the performance. 
Note that the sensor-based closed-loop control is not necessarily better than the open-loop control.

\begin{figure}	
	\centering
	\subfigure{
		\includegraphics[width=0.9\textwidth,keepaspectratio]{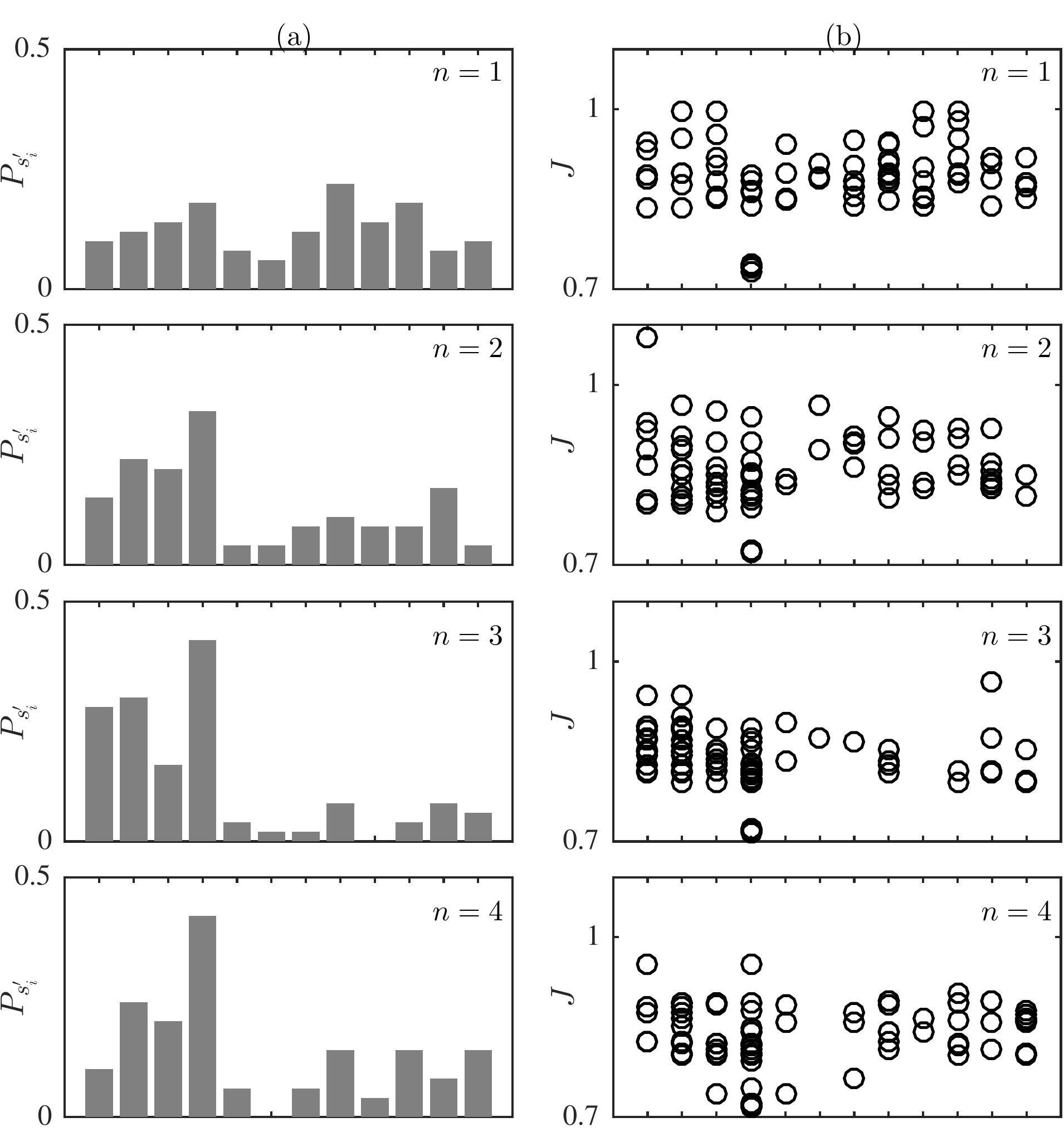}}
	\subfigure{
		\vspace{0mm} 
		\includegraphics[width=0.9\textwidth,keepaspectratio]{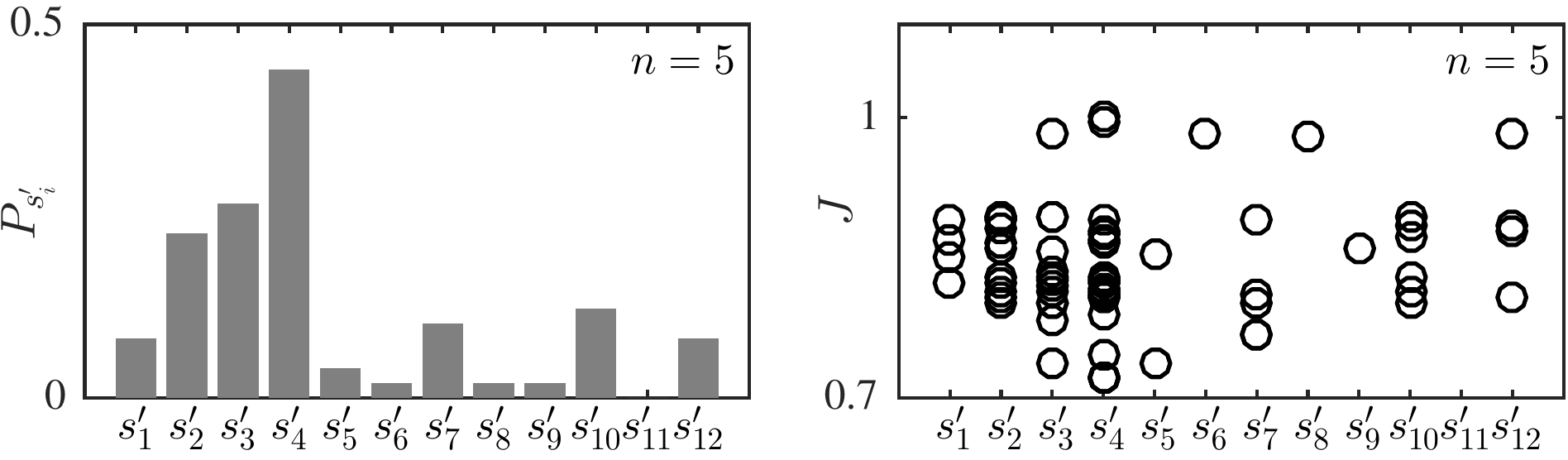}}
	\caption{Convergence of LGPC for single-input multiple-output (SIMO) control.
		For each generation $n=1,...,5$, we represent:
		(\emph{a}) the percentage $P_{s^\prime_i}$ of having $s^\prime_i$ in the expression of the individuals,
		(\emph{b}) the spectrum of $J$-value of individuals which include $s^\prime_i$ in their expression.}
	\label{fig:aSIMO_Distribution_s_J}
\end{figure}

For each generation, figure \ref{fig:aSIMO_Distribution_s_J}(\emph{a}) illustrates 
the percentage $P_{s^\prime_i}$ of having $s^\prime_i$ involved in the individuals, 
and subfigure (\emph{b}) represents the spectra of $J$-values 
of individuals which include  $s^\prime_i$ in their expression. 
Like in \S~\ref{ToC:Quasi periodic forcing}, 
the first generation $n=1$ chooses each sensor signal with comparable percentage.
For a sufficiently large population, all the sensor signals would have nearly equal percentage.  
We observe a minimum $J$-value at $s^\prime_4$ in the first generation. 
The advantage of choosing $s^\prime_4$ is already evident from the second generation. 
Half of the individuals in the following generations 
select $s^\prime_2, s^\prime_3$ and $s^\prime_4$. 
Correspondingly, 
the data points in figure \ref{fig:aSIMO_Distribution_s_J}(\emph{b}) 
represent the progressing move of $J$ from a uniform distribution over all the sensors 
to a concentrated distribution over $s^\prime_2, s^\prime_3$ and $s^\prime_4$. 
Intriguingly, the optimal control law reads 
\begin{equation}
\label{eq:simo_law}
b^{1s}=\mathcal{H}(\tanh(\tanh(s^\prime_4))-0.1).
\end{equation}
Over the 12 sensors, the optimal control law selects only $s^\prime_4$. 
This observation indicates that LGPC provides not only an optimal law but also a sensor selection when multiple sensors are provided to the controller initially.
This optimal law will be physically interpreted in the following sections.


\graphicspath{{./Figures_pdf/}}
\subsubsection{Visualization of SIMO control laws}
\label{sec:aSIMO_visualization}
\begin{figure}
	\centering	
	\begin{overpic}[scale = 0.6]{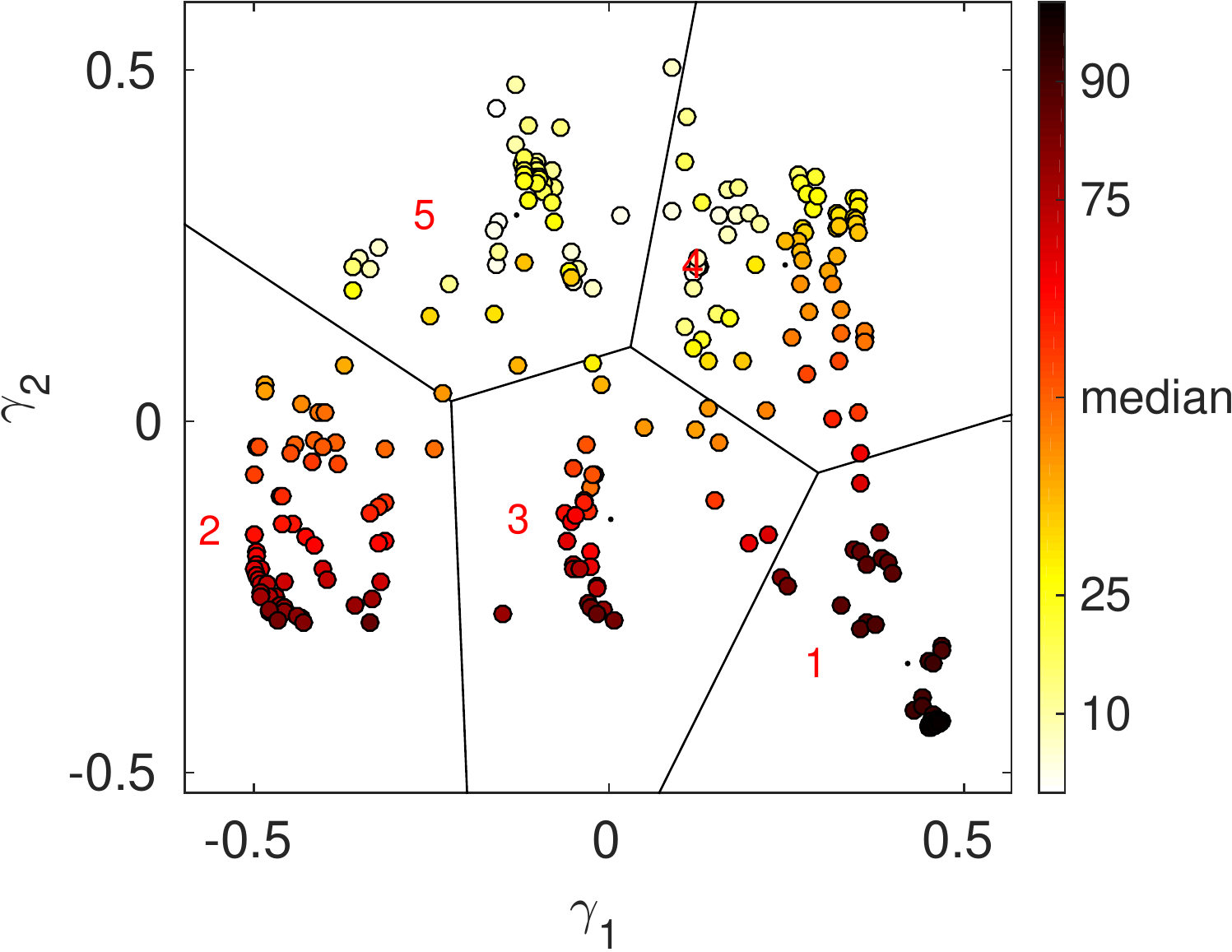}
	 \put(64,23){\fcolorbox{black}{white}{$1$}}
	 \put(15,33){\fcolorbox{black}{white}{$2$}}
	 \put(39,34){\fcolorbox{black}{white}{$3$}}
	 \put(53,55){\fcolorbox{black}{white}{$4$}}
	 \put(33,60){\fcolorbox{black}{white}{$5$}}
	\end{overpic}
	\caption{Visualization of (dis)similarity associated with the entire collection (250 individuals) of the sensor-based SIMO control laws. Each circle represents an individual control law and the distance between two control laws 
	approximates their respective dissimilarity. The colour scheme corresponds to the percentile rank of the control laws with respect to their performance $J$.
	The feature vectors $\{\boldsymbol{\gamma}^l\}_{l=1}^N$ are further analyzed by applying a cluster algorithm.
	The best performing individuals belong to cluster $1$.}
	\label{fig:cMDS_Plot2D_J}
\end{figure}
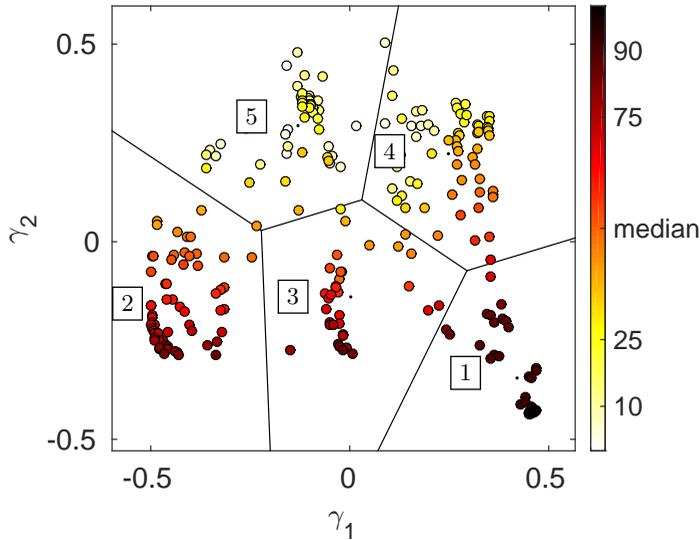

A two-dimensional visualization of control laws is obtained by applying the method described in \S~\ref{sec:Visual}. 
This visualization contributes to get a better understanding of the evolution of control laws. 
The entire collection consisting of $N=M\times G = 50\times5=250$ individuals is considered here. 
The penalization coefficient in the distance matrix \eqref{Eq:DistanceMatrixControlLaws} is chosen to be $\alpha = 3.5$ according to the description in \S~\ref{sec:Visual}. 
CMDS, as explained in \S~\ref{Sec:AppA:CMDS}, yields an ensemble of two-dimensional feature vectors $\{\boldsymbol{\gamma}^l\}_{l=1}^N$, 
with $\boldsymbol{\gamma}^l = (\gamma_1^l,\gamma_2^l)^T$. 
For each individual $l$,
the mutual distances between feature vectors quantify the dissimilarity between different control laws $K^l$.
For further analysis, the ensemble is then partitioned using the k-means clustering algorithm \citep{Lloyd1982ieeetit,Kaiser2014jfm}. 
Mainly five clusters, denoted by $k_c\in\{1,...,5\}$, can be distinguished.
The resulting Voronoi diagram of the clusters is displayed in figure \ref{fig:cMDS_Plot2D_J}.
Each control law is displayed as a circle which is colour-coded by the ordering, here defined in terms of the percentile rank.
For instance, an individual that performs equal or better than $90\%$ of the ensemble of evaluated control laws is said to be at the $90$th percentile rank.

The broad distribution of points over the space illustrates that LGPC has successfully explored a diversity of control laws.
The clusters are ordered according to the mean $J-$value in a cluster. 
Thus, it can be seen from the distribution of $J$ that the control laws in the lower clusters $k_c=1,2,3$ have better performance 
than the upper ones $k_c=4,5$. The top-ranking control laws are located in the cluster $k_c=1$. 
A spectral analysis of the control laws in each cluster shows that 
this clustering partition discriminates their actuation frequency characteristics.
The control laws in the cluster $k_c=1$ exhibit a similar spectrum as that of the optimal actuation $b^{1s}$ shown in figure \ref{fig:aSIMO_LGP}(\emph{c}). 
Their dominant frequency is around $St_{H}=6.9$.
The control laws in its neighbouring cluster $k_c=4$ have the similar dominant frequency as the laws of $k_c=1$. However, they have a larger duty cycle resulting in a different energy distribution in the actuation spectrum. 
The clusters $k_c=3$ and $5$ contain the control laws showing a white noise behaviour with no obvious dominant frequencies. 
The control laws in the upper cluster $k_c=5$ have a larger duty cycle that those of the lower cluster $k_c=3$.
The control laws in the cluster $k_c=2$ possess clearly a high dominant frequency around $St_{H}=8.6$.
It seems that the horizontal coordinate distinguishes the actuation frequencies, whilst the vertical coordinate differentiates the duty cycles.
These observations are consistent with their performance distributions.
By looking into the evolution of points as the generation increases, a global downward shifting can be observed which indicates their convergence to the top-performing individuals.
The visualization provides a simple and revealing picture of the exploration and exploitation characteristics of the control approach, 
inspiring further improvement of the methodology.

\graphicspath{{./Figures_pdf/}}
\subsubsection{Analysis of the optimal control law}
\label{sec:aSIMO_control_laws}
In this subsection, 
we analyse the flow data
with an aim of understanding why
LGPC has chosen the optimal law \eqref{eq:simo_law}. 

We have mentioned previously in \S~\ref{sec:QPF control laws} that better performance is expected 
for a large jet velocity $V_\text{Jet}$ under a high-frequency forcing. 
As we binarize the ON/OFF control command with a Heaviside function, 
an oscillating movement around the threshold of the Heaviside function is responsible to trigger intermittently the actuation. 
High-frequency oscillations lead to high-frequency forcing. 
Therefore, the selected sensors are expected to fulfil three properties. 
First, they should exhibit fluctuations of the unforced baseline 
to provoke the actuation at the very beginning. 
Second, they should highly correlate with the high-frequency forcing 
and yield corresponding fluctuations around the threshold. 
Third, the low-frequency drifts in the sensors 
originating from the motion of the separated bubble or the vortex shedding, 
should not interfere with the high-frequency feedback between actuation and sensing. 
These expected properties guide our analysis of the sensors for insights into the sensor selection.
 
\begin{figure}
	\centering
	\includegraphics[width=0.9\textwidth,keepaspectratio]{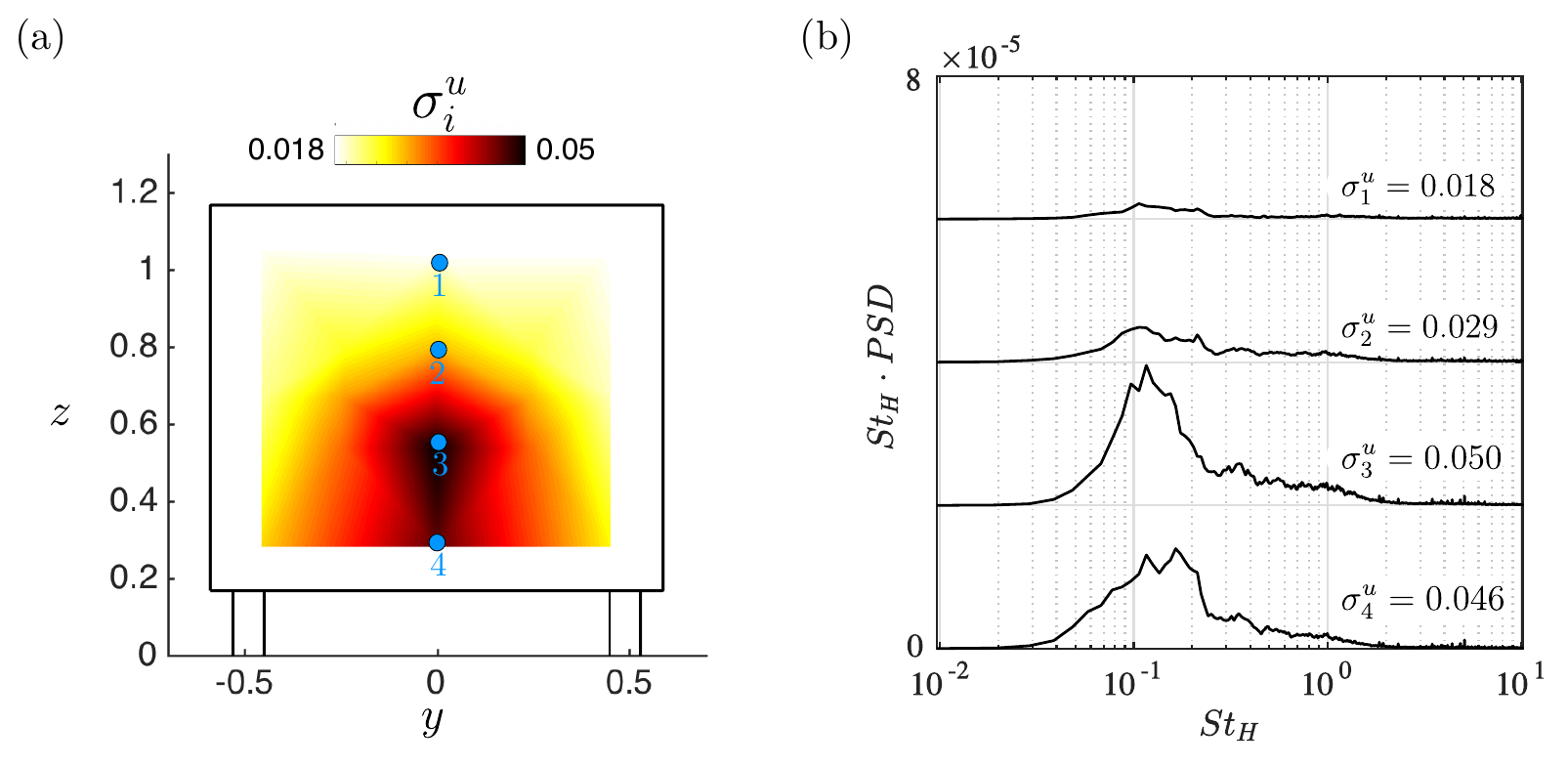}	
	\caption{Characteristics of the sensor fluctuation in the unforced flow. (\emph{a}) Colour map of the standard deviation of the sensor signal $\sigma_i^{u}$, $i=1,\ldots,16$. (\emph{b}) The spectra of sensor signals $s_i^\prime$ located on the symmetry line $y=0$. Values on the vertical axis are shifted for clarity. The levels of standard deviation $\sigma^{u}_{i}$ of $s_i^{\prime}$ $(i=1,\ldots,4)$ are also given in the figure.}
	\label{fig:Nat_sprime_std}
\end{figure}

First, we search for sensors with large fluctuation levels for the unforced flow.
Figure \ref{fig:Nat_sprime_std}(\emph{a}) displays the colour map of the standard deviation $\sigma_i^{u}$ of the sensor signal $s^\prime_i$ $(i=1,\ldots,16)$ for the unforced flow. 
The largest fluctuation level can be observed in the vicinity of the lower edge, 
especially close to the symmetry line $y=0$. 
The spectral analysis is carried out using the signals on this symmetry line. 
The resulting spectra are shown in figure \ref{fig:Nat_sprime_std}(\emph{b}) 
with a vertical shift for clarity. 
Clearly, $s_3^\prime$ and $s_4^\prime$ feature a larger fluctuation level than the others. 
Intriguingly, their positions are very close to the attachment point on the base surface of two trapped vortices (see figure \ref{fig:PIV_wake}(\emph{b})). 
The vortex shedding mode around $St_{H}=0.2$ is hardly be seen in this figure. 
The important energy content around $St_{H}=0.1$ 
in each sensor indicates a global motion of the separation bubble. 
The energy reaches its maximum in sensor $s_3^{\prime}$. 
Based on this observation, 
we assume that $s_3^\prime$ and $s_4^\prime$ could be the desired candidate sensors in LGPC. 
\graphicspath{{./Figures_pdf/}}

The next analysis concerns the forced flow.
Characteristic features of high-frequency forcing
are the large time delay ($\tau_d=\SI{5}{\milli\second}$) 
from actuation to sensing and 
the high level of correlation between actuation and sensing 
at the frequency of actuation 
(see  \S~\ref{sec:QPF control laws}).
The time delay roughly corresponds to two periods of optimal periodic forcing.
For closed-loop control, these features indicate 
that the actuation pulse will be felt by the sensors after time $\tau_d$, 
and this oscillation will trigger in real-time another actuation pulse.
In other words,  an actuation pulse is triggered by the effect of previous pulses. 
Once some stochastic flow perturbations produce a high-frequency sensor oscillation 
around the right threshold, 
the system would self-sustain the high frequency forcing. 
This explains why the optimal feedback law yields such highly periodic dynamics. 
As presented in \S~\ref{sec:QPF control laws}, 
almost all the sensors have a high correlation with the forcing. 
We assume that both $s_3^\prime$ and $s_4^\prime$ have the capability to capture and amplify the perturbation created by the actuation and feed it back to maintain the forcing. 

\begin{figure}
	\centering
	\includegraphics[width=1\textwidth,keepaspectratio]{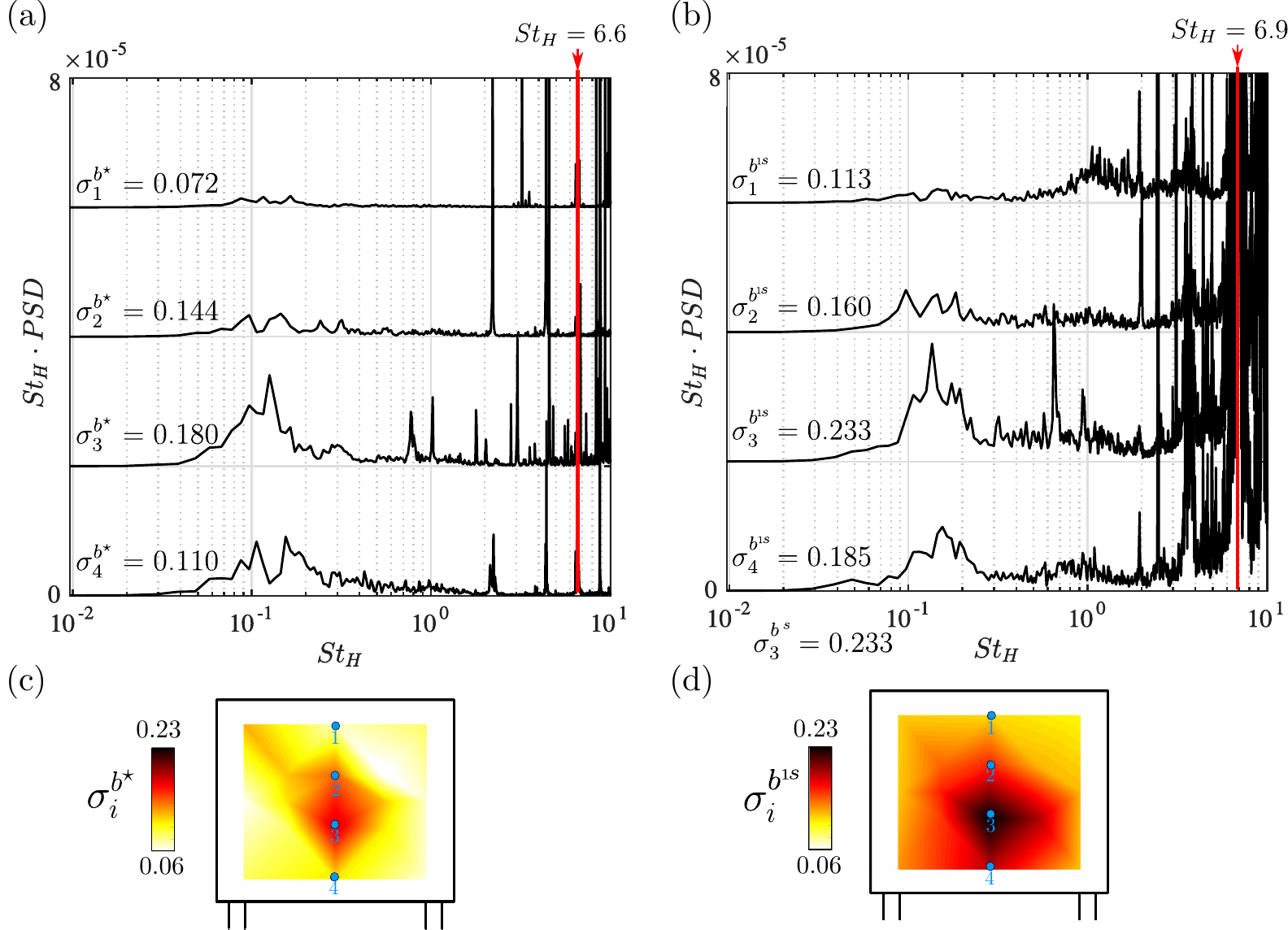}	
	\caption{Characteristics of the sensor fluctuation in forced flows for (\emph{a,c}) the optimal SIPF $b^{\star}$ and (\emph{b,d}) the optimal SIMO control $b^{1s}$. (\emph{a,b}) Spectra of sensor signals on the symmetry line $y=0$. The dashed red line indicates the dominant forcing frequency. (\emph{c,d}) Colour maps of the standard deviation of the sensor signals.}
	\label{fig:PF_aSIMO_sensor_spectra}
\end{figure}

We now focus on the low-frequency spectrum of the sensor signals under the forcing. 
Figure \ref{fig:PF_aSIMO_sensor_spectra} shows the spectra of the sensors on the symmetry line ($y=0$) under the optimal SIPF $b^{\star}$ (\emph{a}) and the optimal SIMO forcing $b^{1s}$ (\emph{b}). 
The colour map of the standard deviation of the sensor signal $s^\prime_i$ $(i=1,\ldots,16)$ is presented in figure \ref{fig:PF_aSIMO_sensor_spectra}(\emph{c}) and (\emph{d}) for $b^{\star}$ and $b^{1s}$, respectively.
The forcing frequency, highlighted by the red line, is felt by all sensors  in agreement with the previous results in \S~\ref{sec:QPF control laws}. 
An important observation is that the low-frequency drift of the unforced and forced flow are similar.
This indicates that high-frequency forcing does not noticeably modify the dynamics of the large-scale structures as felt by the sensors. 
The maximum energy at low frequencies is reproducibly found for $s_3^\prime$. 
However, this property is a disadvantage for  
selecting $s_3^\prime$ as feedback sensor, due to the third postulated sensor property. 
In addition, 
the spectrum of $s_4^\prime$  is less noisy in the high-frequency range for open- and closed-loop forcing
as compared to $s_3^\prime$.
All these considerations lead naturally to the selection of $s_4^\prime$ for feedback.

\graphicspath{{./Figures_pdf/}}

In summary, $s_4^\prime$ captures, on the one hand, 
strong enough dynamics in the unforced flow to trigger the feedback cycle 
and, on the other hand, small enough low-frequency dynamics in the forced flow 
to maintain the fluctuations around the trigger threshold. 
Given these conditions, $s_4^\prime$ is capable to create a nearly periodic high-frequency forcing 
and it self-adapts to converge to the optimal periodic forcing. 
The stochastic fluctuations relate to small-scale structures.
These fluctuations add to a noisy spectrum in the closed-loop control. 
In time domain, this indicates that there exist a variety of pulse durations $t_\text{Pulse}$ 
and intermittent quiet time $t_\text{Int}$ in the actuation command. 
In light of the analysis of \S~\ref{sec:QPF control laws}, 
this variety may influence the instantaneous curvature of the shear layer 
and degrade globally the control performance.

\subsection{Morlet filtering of sensor signals}
\label{sec:MI_LGPC}
\begin{figure}
	\centering
	\includegraphics[width=0.9\textwidth,keepaspectratio]{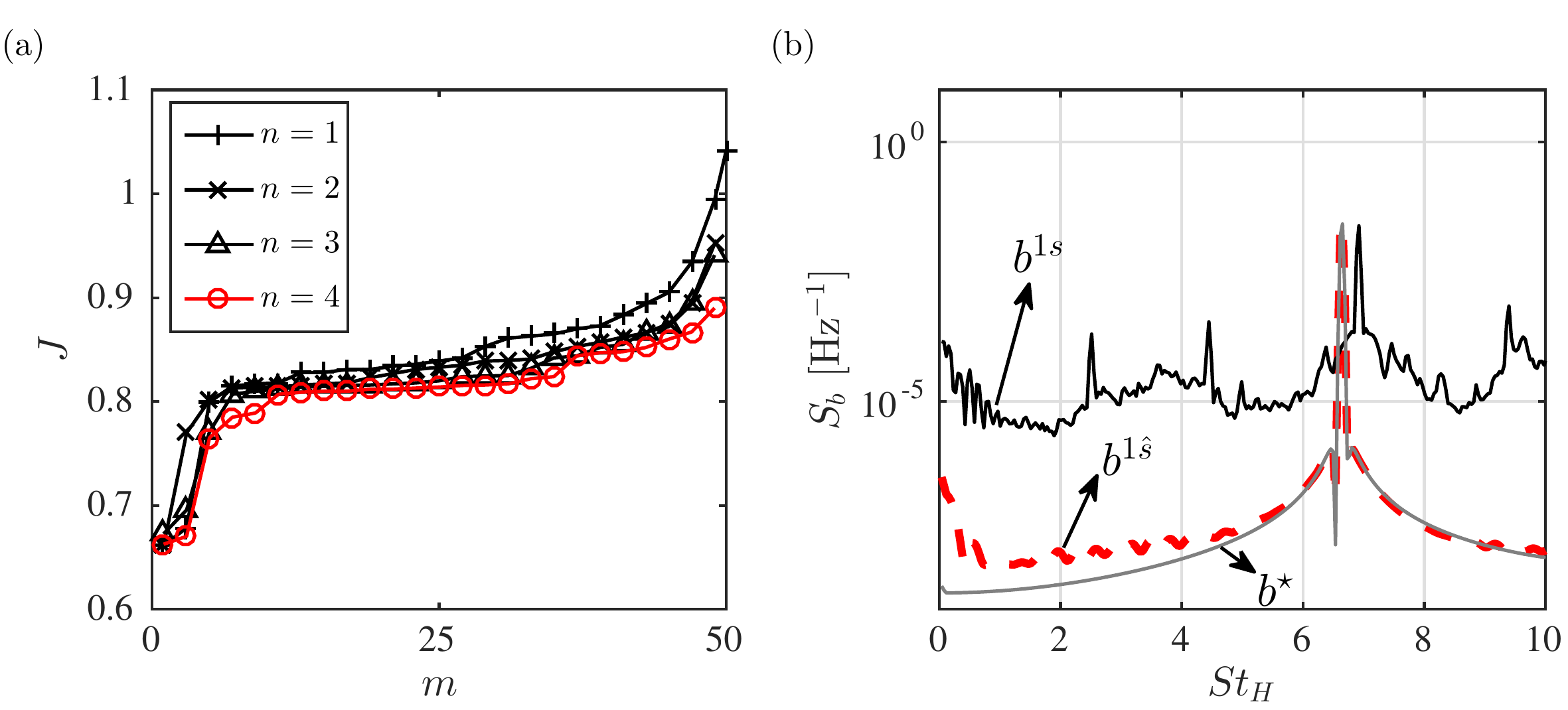}	
	\caption{Results of LGPC for the single-input single-output control with Morlet filtering. (\emph{a}) Evolution of the cost function $J$ versus the individuals for four generations $n=1,\ldots,4$. (\emph{b}) Power spectral density $S_{b}$ of the optimal SISO control law $b^{1\hat{s}}$. The spectra of the optimal SIPF $b^{\star}$ and the optimal SIMO control law $b^{1s}$ are also given for comparison.}
	\label{fig:aSIMOMMF_LGP}
\end{figure}

\begin{figure}
	\centering
	\includegraphics[width=0.4\textwidth,keepaspectratio]{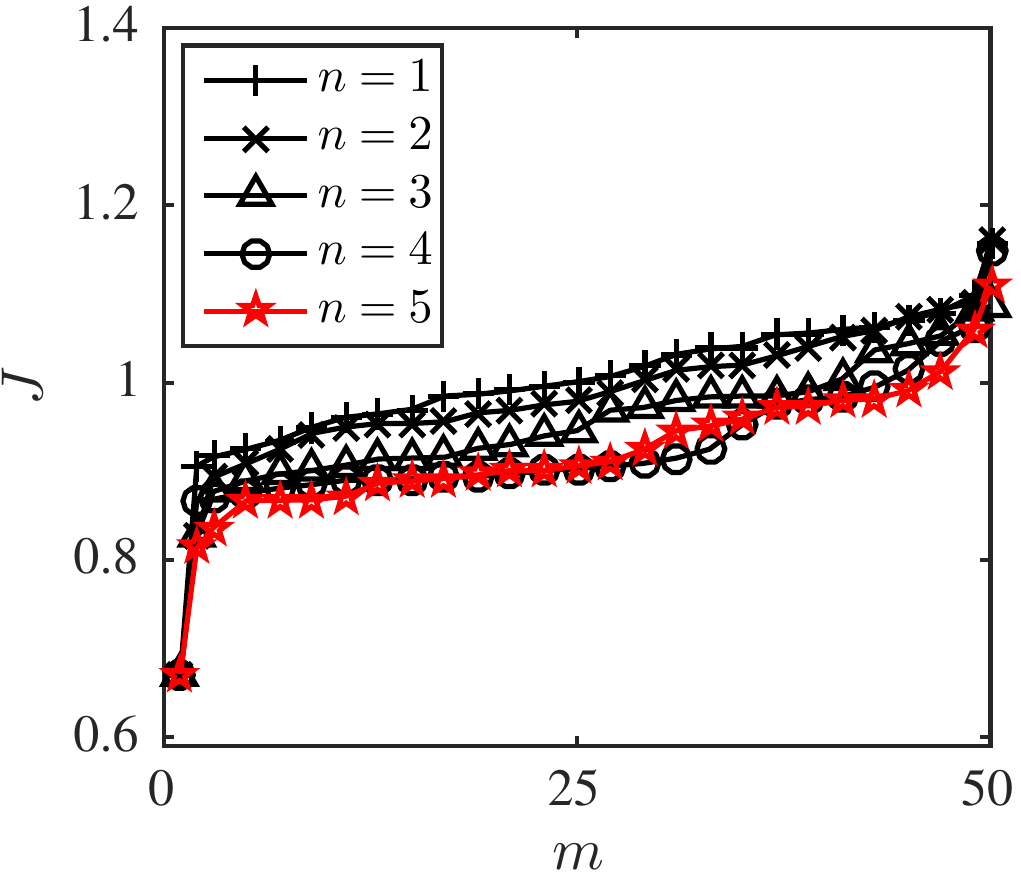}	
	\caption{Multiple-input single-output control. Evolution of the cost $J$ for the individuals.}
	\label{fig:MISO_J_ind}
\end{figure}

In this section, we explore the potential benefits
of extracting  frequencies of interest in the sensor signals by applying a specific filter.
A Morlet wavelet Filter (MF) as described in \S~\ref{Sec:AppB:CLMF} is particularly suited to this task.
Only the filtered time-history feedback of the fourth sensor is considered,
resulting in SISO (Single-Input Single-Output) control when the four actuators are driven in unison, and
MISO (Multiple-Input Single-Output) control when the four actuators are independent.
For SISO, the results are shown in figure \ref{fig:aSIMOMMF_LGP}. Only four generations are conducted, and the optimal control law has been found from $n=1$. The optimal control law $b^{1\hat{s}}=\mathcal{H}(\tanh{(\hat{s}_4})-0.13)$ contains only one sensor $\hat{s}_4$ over the 7 sensors defined in \S~\ref{Sec:AppB:CLMF}. We remind that $\hat{s}_4$ is the filtered signal of $s$ through the wavelet $\psi_4$ with the centred frequency at $St_{H_c}=6.5$. The spectra of the $b^{1s}$, $b^{1\hat{s}}$ and $b^{\star}$ are shown in figure \ref{fig:aSIMOMMF_LGP} (\emph{b}). $b^{1\hat{s}}$ shows a much smoother spectrum than $b^{1s}$ indicating that the noisy fluctuations are filtered out by the MF. As a result, $b^{1\hat{s}}$ leads to approximately a single-frequency forcing which is the same as the optimal SIPF $b^{\star}$.       
\graphicspath{{./Figures_pdf/}}
For MISO,  
the evolution of $J$ over five generations is shown in figure \ref{fig:MISO_J_ind}. 
We include the optimal SISO individual $b^{1\hat{s}}$ into the first generation of LGPC. 
$b^{1\hat{s}}$ remains unchallenged after 5 generations. 
The abrupt decline of $J$ between the first and second individual evidences 
the longer learning time of  MISO as compared to SISO. 

\graphicspath{{./Figures_pdf/}}

\subsection{Generalized non-autonomous control}
\label{sec:other_cases}
We explore a generalized non-autonomous control by comprising the sensors feedback $\vec{s}$ and the harmonic functions as the inputs of control law. 
In this case, the control law is expressed by $b=K(\vec{s},t)$, 
where $t$ represents the time. 
In the present study, we include only the harmonic function $h^\star=\sin(2\pi f^\star t)$ as a supplementary input of control laws, where $f^\star$ is the frequency of the optimal SIPF $b^\star$. 
Three configurations are studied: SIMO, MIMO and SISO. 
Only one or two generations are performed 
because LGPC quickly converges to the optimal results observed previously. 
For SIMO and MIMO, 
LGPC finds the optimal SIPF $b^{\star}$ as the top-performing individual. 
In SISO, the optimal control law $b^{1\hat{s}}$ obtained in \S~\ref{sec:MI_LGPC} 
and the optimal SIPF $b^{\star}$ win the evolution. 
A further study by comprising more harmonic functions is envisaged.

\section{Conclusions}
\label{ToC:Conclusions}
We present the first application 
of genetic programming control (GPC) for aerodynamic drag reduction.
The well-investigated blunt-edged Ahmed body has been chosen as benchmark configuration.
The flow is manipulated by 4 jet actuators with Coanda surface deflectors
at all trailing edge and is monitored by 16 pressure sensors at the rear side, 12 being used for the feedback.
The control performance is graded by the averaged base-pressure coefficient.
This performance measure was shown to be strongly correlated with the drag
for periodic forcing from low to high actuation frequencies.

In earlier publications \citep{Barros2015phd,Barros2016jfm},
high-frequency periodic forcing reduces the drag by 20\%.
Evidently the perceivable space of control laws is significantly larger.
By definition, multi-frequency forcing may generalize the best periodic actuation
and should hence be better --- or at least not worse. 
So far, this was not explored for the Ahmed body drag reduction. 
There are reported benefits for multi-frequency forcing in other flow control configurations,
for instance pressure recovery in a diffuser \citep{Narayanan1999utrc}.
Sensor-based feedback has been shown to outperform periodic forcing
for drag reduction of a D-shaped body \citep{Pastoor2008jfm}.
A very general method for sensor-based feedback is provided by 
genetic programming control (GPC)
\citep{Gautier2015jfm,Debien2016ef,Parezanovic2016jfm}.
Yet, the advantages of filtering out noise has hardly been explored in GPC \citep{Duriez2016book}.
In addition, there is no a priori reason 
why sensor-based feedback should be better than optimized forcing. 
In fact, a mixing layer control study \citep{Parezanovic2016jfm} shows
that optimized periodic forcing may be better or worse than sensor-feedback
depending on the location of the sensors and the definition of the cost functional.
However, closed-loop control laws may be formulated 
in a manner to include the optimal open-loop control.
For instance, the optimal actuation command 
may be employed as additional artificial sensor signal.
The number of sensors and actuators contribute 
the challenge to explore the full search space of control laws.
For example, it is easy to tune the frequency of a single actuator,
but not to optimize independent multi-frequency forcing in  4 independent actuators.

In this study, we have significantly 
enlarged the search space of control laws 
by incorporating the above mentioned successful strategies for other configurations.
The reference of this study was an optimized periodic forcing
$b = K \left ( \sin (2\pi f^{\star} t) \right)$.
This open-loop actuation is generalized
by including harmonic functions $h_i = \sin (2\pi f_i t)$
for 9 frequencies $f_i$, $i=1,\ldots,9$,
including the optimal one $f^\star $. 
These functions are comprised in the time-dependent vector 
$\vec{h}(t)=(h_1,\ldots,h_9)^T(t)$.
The resulting actuation law reads 
\begin{equation}
 b_j= K \left ( \vec{h} \right ) 
= K\left( \sin (2\pi f_1 t), \ldots, \sin(2\pi f_9  t) \right),
\quad j=1, \ldots, 4
\label{eq:SIMFF}
\end{equation}
where $j$ represents the $j$th control. 
With $b_1=b_2=b_3=b_4$, all actuators are driven in single-input mode, \emph{i.e.} in unison.
Another departure point of this study 
is sensor feedback from GPC,
$b_j = K (\vec{s}^{\prime}) = K \left( s^{\prime}_1, \ldots, s^{\prime}_{12} \right) $,
with one actuation input and all 12 sensor signals.
As first generalization, 
we want to take full advantage of the multi-input capability,
\emph{i.e.} of driving the four actuators independently:
\begin{equation}
b_j=  K_j (\vec{s}^\prime) = K_j \left(s^{\prime}_1, \ldots, s^{\prime}_{12} \right) , \quad j=1,\ldots,4.
\label{eq:SIMO}
\end{equation}
Following a recommendation of Wahde (2013, personal communication),
\emph{linear} genetic programming (LGP) is employed as regression technique.
Arguably, LGP is more suited for multiple inputs 
than tree-based genetic programming (TGP).
Certainly, 
LGP regression is  much easier to code than TGP.
In principle, LGP can represent any TGP-based law and the other way round.
Second, we include Morlet-filtered sensor signals ${\hat s}_{i,c_k}$,
where $i$ represents the sensor and $c_k$ 
refers to one of 5 filter frequencies $f_{c_k}$, $c_k=1,\ldots,5$.
These frequencies are selected from the 9 mentioned actuation frequencies.
All filtered signals are incorporated in $\hat{\vec{s}}=(\hat{\vec{s}}_{1}, \ldots, \hat{\vec{s}}_{5})^T$, where
$\hat{\vec{s}}_{c_k}=(\hat{s}_{1,c_k}, \ldots, \hat{s}_{12,c_k})^T$, $c_k=1,\ldots,5$.
The resulting control law 
\begin{equation}
b_j
= K_j\left (\vec{s}^\prime,\hat{\vec{s}} \right ), \quad j=1,\ldots,4
\label{eq:generalcase1}
\end{equation}
generalizes the instantaneous sensor-based feedback
by including sensor history.
Note that each control law $K_j$  has about 72 arguments.
Both, multi-frequency forcing and feedback with sensor history,
are included in the non-autonomous feedback law
\begin{equation}
b_j 
= K_j \left (\vec{s}^\prime,\hat{\vec{s}}, \vec{h} \right ) 
= K_j \left(\vec{s}^\prime,\hat{\vec{s}},
\sin(2 \pi f_1 t), \ldots, \sin( 2\pi f_9  t) \right), \quad j=1,\ldots,4.
\label{eq:generalcase2}
\end{equation}
Its  vector form reads
$\vec{b} = \vec{K}  \left (\vec{s}^\prime,\hat{\vec{s}}, \vec{h} \right )$
with $\vec{b}=(b_1,\ldots,b_4)^T$ and $\vec{K}=(K_1,\ldots,K_4)^T$.

Note that this control law has 4 functions with 81 arguments each.
We did not attempt to solve this most general regression problem.
Instead, we have navigated through the decision space incrementally,
to explore the effect of increasing the number of inputs, 
of including filtered signals 
and of including harmonic inputs, individually.

\graphicspath{{./Figures_pdf/}}
\begin{figure}	
	\centering
	\includegraphics[width=0.9\textwidth,keepaspectratio]{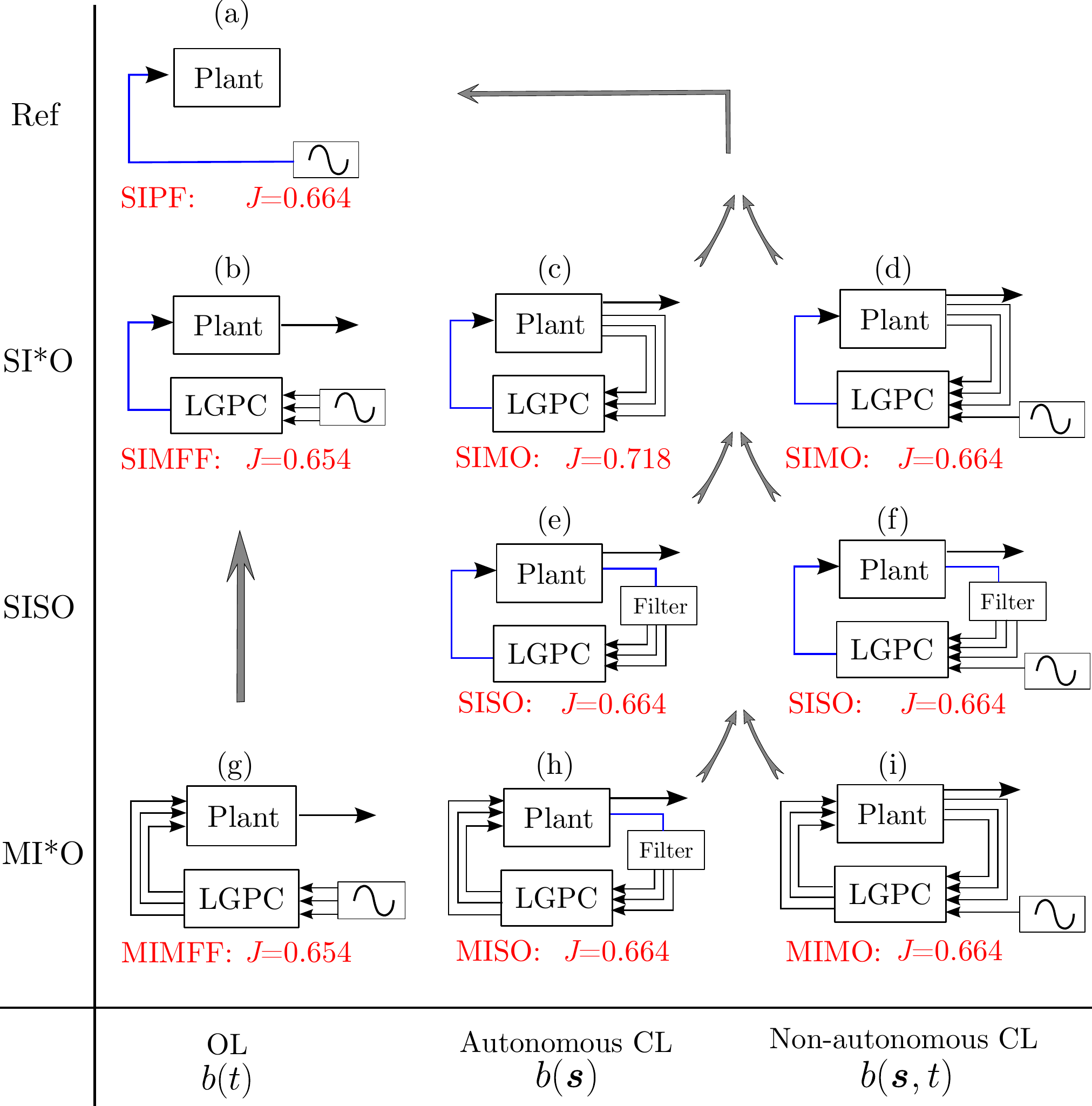}	
	\caption{Synthesis of the LGPC for the investigated classes of control laws. In vertical direction, the control laws are classified by the number of inputs and outputs with respect to the plant. In horizontal direction, the control laws are classified by their arguments. 
	Here, OL and CL refer to open-loop and closed-loop control, respectively.
	A single arrow represents a  single input or output. Three parallel arrows represent multiple-input/output. The inclusion of the best frequency is indicated by an arrow from the sine generator (box with `$\sim$' ). Multiple frequencies are depicted by 3 parallel arrows. The abbreviations and performances (cost value $J$) for each control law are given under each diagram. For further details, see text.}
	\label{fig:Synthesis}
\end{figure}

Figure \ref{fig:Synthesis} summarizes all the investigated classes of control laws.
Single-input control was explored first.
Single-input periodic forcing (SIPF), presented in figure \ref{fig:Synthesis}(\emph{a}), was studied to build a reference for all control designs.
The optimal frequency $St^\star_{H}=6.6$ with $DC^\star$=33\% leads to about 33\% base pressure recovery associated with 22\% drag reduction.
The single-input multi-frequency forcing (SIMFF) generalizes SIMO using LGPC (\ref{eq:SIMFF}), as presented in figure \ref{fig:Synthesis}(\emph{b}).
LGPC yields a two-frequency forcing which outperforms the optimal SIPF reference. 
This control has been identified by testing only 200 individuals in less than one hour.
This testing time is less than employed for finding the best frequency and duty cycle for the periodic reference
with an exhausting parameter sweep.

After exploring the open-loop forcing space, we turn to the sensor-based feedback control space.
With no \textit{a priori} knowledge about the sensors, multiple sensors are explored giving a SIMO control (\ref{eq:SIMO}), as presented in figure \ref{fig:Synthesis}(\emph{c}).
The corresponding closed-loop actuation emulates the optimal high-frequency SIPF.
As to the authors' knowledge this is the first realization of a direct sensor feedback control at a high frequency.
In addition, LGPC reproducibly selects only one sensor near the centre of bottom edge in the optimal control law. 
Thus, LGPC provides not only an optimal actuation but also a sensor optimization for a general class of control laws.
More importantly, LGPC finds a SISO control as the optimal control law in a SIMO framework.
The observation guided us to explore SISO control with the optimal sensor.
We also include the Morlet-filtered signals of the sensor in the control law to include history information for SISO, as depicted in figure \ref{fig:Synthesis}(\emph{e}).
The resulting control law outperforms the optimal SIMO control.

The results of LGPC in SIMO and SISO show both the trend of convergence to the reference SIPF. 
We included, as a first test case, the harmonic function with the frequency in the reference SIPF as an additional artificial sensor for the feedback control laws, as presented in figure \ref{fig:Synthesis}(\emph{d}) for SIMO and (\emph{f}) for SISO.
In this case, LGPC yields a generalized non-autonomous closed-loop control, and automatically selects open-, closed-loop or combination thereof depending on their performance.
In the present study, LGPC has selected in both cases the optimal SIPF as the optimal control law.

Multiple-input control has also been explored by LGPC for MIMFF, MISO and MIMO, as presented in figure \ref{fig:Synthesis}(\emph{g}-\emph{i}).
The optimal respective single-input control law was inserted in the first generation to accelerate convergence. LGPC with multiple inputs did not improve the performance for the best single-input law.
Convergence of LGPC was much slower without this preparation of the first generation. One reason may be the larger search space of control laws.

The approach can be applied to virtually any control problems with a MIMO plant experimentally and numerically. 
For instance, \citet{Grandemange2013jfm} have shown the existence of bi-stability in the wake of a blunt-edged body which may lead to induced drag. 
LGPC can be applied to suppress this bi-stability using different actuations on the left and right edge. 
\citet{Pastoor2008jfm} has applied a robust feedback controller on a D-shape body operating with an actuation frequency which is smaller than the natural vortex shedding periodic motion and obtained 15\% drag reduction. 
The results of the D-shape body shows that significant drag reduction may be obtained by LGPC if the sensors are carefully low-pass filtered. 
For numerical application, e.g. \emph{large eddy simulation}, our methodology will be much more time-consuming \citep{krajnovia2005jfea,krajnovia2005jfeb}.

The model-free foundation of LGPC can be considered as a weakness
since understanding is not probed in form of a control-oriented model.
However, model-based control will be affected by model errors.
In addition, the actuation effect on turbulent flow is quite a challenge to model.
In particular, frequency crosstalk in nonlinear coherent structure interactions typically defies any attempts towards a low-dimensional control-oriented model.
Hence, a model-free foundation removes a significant source of errors and limitations
which - more often than not - hardwires less effective actuation mechanics.
The authors currently improve the LGPC methodology, and pursue Ahmed body experiments improving MIMO control for drag reduction with a yaw angle,
for mitigating the asymmetry by bi-stability 
and for exploring low-frequency actuation. 

\begin{acknowledgments}
\section*{Acknowledgements}
The authors acknowledge the great support during the experiment by J.-M. Breux, J. Laumonier, P. Braud
and R. Bellanger. The thesis of RL is financially supported by PSA Peugeot-Citro\"{e}n 
in the context of OpenLab Fluidics between Peugeot-Citro\"{e}n and
Institute Pprime (Fluidics$ @ $poitiers). 
The authors acknowledge the funding and excellent working conditions 
of the Collaborative Research Center (CRC 880) 'Fundamentals of High Lift for Future Civil Aircraft' funded by the German Research Foundation (DFG)
and hosted at the Technical University of Braunschweig, Germany, and of the former Chair of Excellence
'Closed-loop control of turbulent shear flows 
using reduced-order models' (TUCOROM, ANR-10-CHEX-0015)
supported by the French Agence Nationale de la Recherche (ANR)
and hosted by Institute Pprime.
We would also like to acknowledge the support of the ANR SepaCoDe (ANR-11-BS09-018) and the ONERA INTACOO grants.


We appreciate valuable stimulating discussions
with:
Markus Abel,
Diogo Barros,
Steven Brunton,
Sini\v{s}a Krajnovi\'{c},
Vladimir Parezanovi\'{c},
Rolf Radespiel,
Peter Scholz,
Richard Semaan,
Andreas Spohn and
Mattias Wahde.

\end{acknowledgments}

%
%

\bibliographystyle{bst/jfm}

\bibliography{./Bib/biblio_used,./Bib/biblio_bernd}


\appendix
\section{Classical multidimensional scaling (CMDS)}
\label{Sec:AppA:CMDS}
Classical multidimensional scaling (CMDS) is employed to visualize the similarity of control laws (see \S~\ref{ToC:LGPC}).
CMDS aims to find a low-dimensional representation of points $\boldsymbol\gamma^l$, $l=1,\ldots,N$, 
such that the average error between the distances between points $\boldsymbol\gamma^l$ and the elements of a given distance matrix $\boldsymbol{\mathsfbi{D}}$, 
here emulating the distance between the time series of different control laws, is minimal.

In order to find a unique solution to CMDS, 
we assume that $\boldsymbol{\Gamma}=[\boldsymbol{\gamma}^1\quad \boldsymbol{\gamma}^2\quad \ldots\quad\boldsymbol{\gamma}^N]$ with $\boldsymbol{\gamma}^1,\ldots,\boldsymbol{\gamma}^N\in\mathbb{R}^r$ is centered, 
i.e.\ $\boldsymbol{\Gamma}$ is a mean-corrected matrix with $1/N\,\sum_{l=1}^N\, \boldsymbol{\gamma}^l = [0 \ldots 0]^T$.
Rather than directly finding $\boldsymbol{\Gamma}$, we search for the Gram matrix 
$\boldsymbol{\mathsfbi{B}}=\boldsymbol{\Gamma}^T\boldsymbol{\Gamma}$ that is real, symmetric and positive semi-definite.
Since $\boldsymbol{\Gamma}$ is assumed to be centered, the Gram matrix is the Euclidean inner product, and we have 
$\mathsfi{D}_{lm}^2=\vert\vert \boldsymbol{\gamma}^l-\boldsymbol{\gamma}^m\vert\vert_2^2=\mathsfi{B}_{ll} + \mathsfi{B}_{mm} - 2\,\mathsfi{B}_{lm}$.
In the first step of the classical scaling algorithm, the matrix $\boldsymbol{\mathsfbi{D}}_2$ of elements $\left(\mathsfi{D}_2\right)_{lm} = -\frac{1}{2}\mathsfi{D}_{lm}^2$ is constructed.
Then, we form the "doubly centered" matrix  $\boldsymbol{\mathsfbi{B}} = \boldsymbol{\mathsfbi{C}}\boldsymbol{\mathsfbi{D}}_2\boldsymbol{\mathsfbi{C}}$, where $\boldsymbol{\mathsfbi{C}}=\boldsymbol{\mathsfbi{I}}_N-N^{-1}\boldsymbol{\mathsfbi{J}}_N$ with $\boldsymbol{\mathsfbi{I}}_N$ the identity matrix of size $N$ and $\boldsymbol{\mathsfbi{J}}_N$ an $N\times N$ matrix of ones. The term "doubly centered" refers to the subtraction of the row as well as the column mean.
Let the eigendecomposition of $\boldsymbol{\mathsfbi{B}}$ be $\boldsymbol{\mathsfbi{B}} = \boldsymbol{\mathsfbi{V}}\boldsymbol{\mathsfbi{\Lambda}}\boldsymbol{\mathsfbi{V}}^T$ 
where $\boldsymbol{\mathsfbi{\Lambda}}$ is a diagonal matrix with ordered eigenvalues
$\lambda_1\geq \lambda_2 \geq\ldots\geq\lambda_N\geq 0$
and $\boldsymbol{V}$ contains the eigenvectors as columns.
Then $\boldsymbol{\Gamma}$ can be recovered from
\begin{equation}
\boldsymbol{\Gamma} = \boldsymbol{\mathsfbi{\Lambda}}^{\frac{1}{2}}\boldsymbol{\mathsfbi{V}}^T.
\end{equation}
Having only the distance matrix, the resulting representation is only defined up 
to a translation, a rotation, and reflections of the axes.
If the distance matrix is computed using the Euclidean distance and all eigenvalues are non-negative,
$\boldsymbol{\Gamma}$ can be recovered.
If $r<N$, there exist $N-r$ zero eigenvalues, 
in which case a low-dimensional subspace can be found in which the presentation of $\boldsymbol{\Gamma}$ would be exact.
For other distance metrics, the distances of the presentation found by CMDS is an approximation to the true distances.
In this case, some eigenvalues may be negative and only the positive eigenvalues and their associated eigenvectors 
are considered to determine an approximative representation of $\boldsymbol{\Gamma}$.
Note that for the Euclidean distance metric, CMDS is closely related to a principal component analysis (PCA)
commonly used to find a low-dimensional subspace. 
While CMDS, and multi-dimensional scaling generally, uses a distance matrix as input, PCA is based on a data matrix.
A distance matrix $\boldsymbol{\mathsfbi{D}}$ can be directly computed for the centered matrix $\boldsymbol{\Gamma}$. 
If the Euclidean distance is employed for computing the distances, 
the result from applying CMDS to $\boldsymbol{\mathsfbi{D}}$ corresponds to the result 
from applying PCA to $\boldsymbol{\Gamma}$.
A proof can be found in \cite{Mardia1979book}.
The quality of the representation is typically measured by 
$\sum_{l=1}^r\,\lambda_l/\sum_{l=1}^{N-1}\,\lambda_l$, 
and more generally if $\boldsymbol{\mathsfbi{B}}$ is not positive semi-definite using
$\sum_{l=1}^r\,\lambda_l/\sum_{\lambda>0}\,\lambda_l$.
\section{Feedback control using Morlet filters}
\label{Sec:AppB:CLMF}
\begin{figure}
	\centering
	\includegraphics[scale = 0.5]{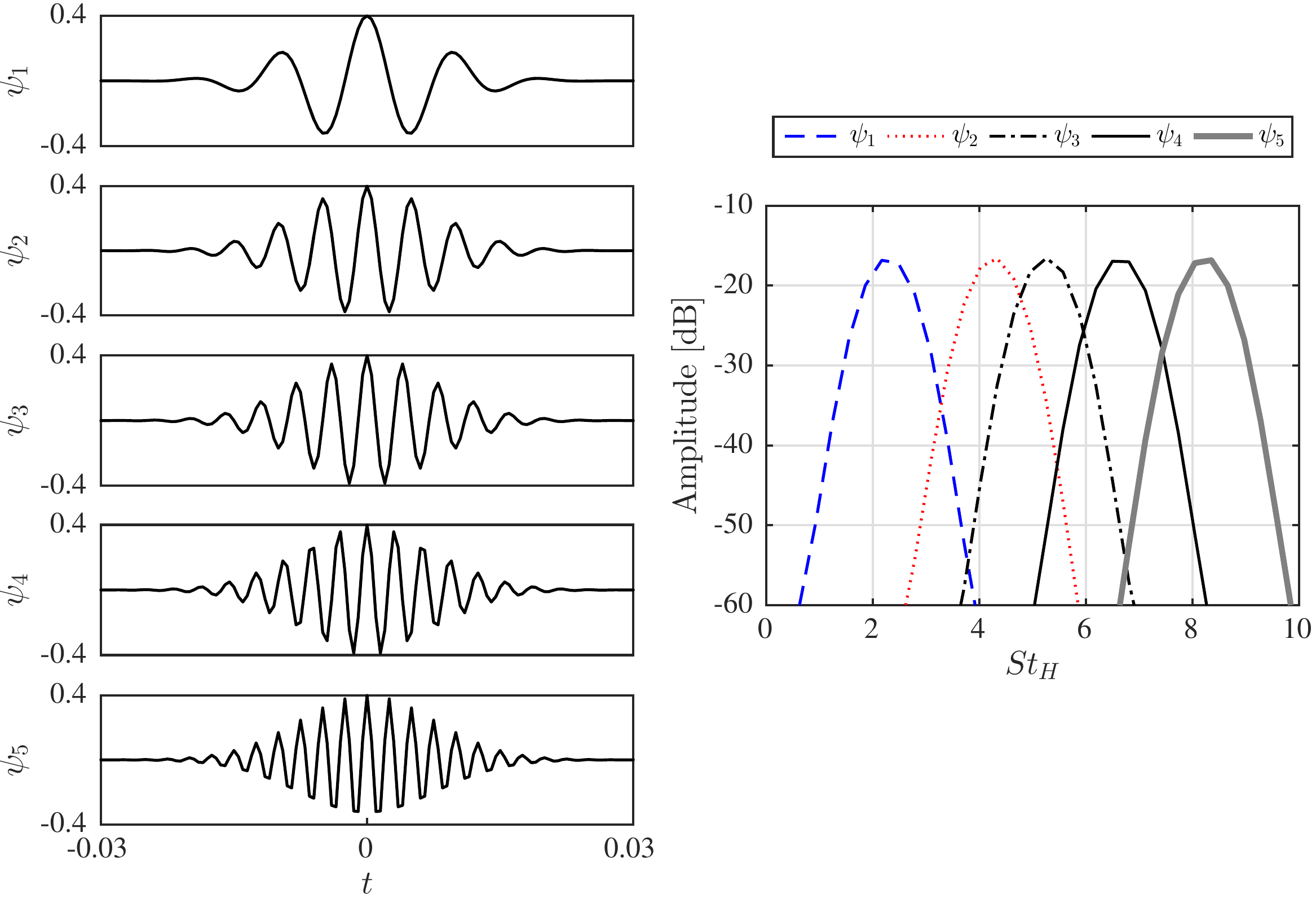}
	\caption{Morlet wavelets in time domain (left) and frequency domain (right).}
	\label{fig:MF1_5}	
\end{figure}

In this section, we describe the use of Morlet wavelet filter (MF) to extract frequencies of interest 
in the sensor signals.
In time domain, the Morlet wavelet $\psi$ is a cosine function modulated by a Gaussian envelope.
It is then defined for a frequency $f_{c}$ as:
\begin{equation}
\psi(t)=\dfrac{1}{\sqrt{2\pi}\sigma}\exp(-\frac{t^2}{2\sigma^2})\cos(2\pi f_{c}t).
\label{eq:MF}
\end{equation}
In frequency domain, MF is a band-pass filter which attenuates the undesired frequencies outside the range $[f_{c}-\lambda/2, f_{c}+\lambda/2]$, where $\lambda$ represents the bandwidth which is governed by the parameter $\sigma$.
In our applications, only the fourth sensor $s_4$ identified for the optimal SIMO control (see \S~\ref{sec:aSIMO}) is chosen as the output of the plant, resulting in SISO (Single-Input Single-Output) system. 
To avoid the confusion, we denote the fourth sensor $s_4$ as $s$ and its fluctuation $s^\prime_4$ as $s^\prime$. 
The sensor $\vec{s}$ in the feedback control law $b=K(\vec{s})$ is defined as 
$\vec{s}=[\hat{s},\ldots,\hat{s}_5, \overline{s},s^\prime]$, where
\begin{equation}
\begin{split} 
\hat{s}_i(t)&=\int_{0}^{\tau_P}\psi_i(\tilde{t})s^\prime(t+\tilde{t}-\tau_P)\text{d}\tilde{t},\quad i=\{1,...,5\}\\
\overline{s}(t)&=\frac{1}{\tau_P}\int_{t-\tau_P}^{t}s(t)\text{d}t\\
s^\prime(t)&=s-\overline{s}(t).
\end{split}
\label{eq:morlet_wavelet1}
\end{equation}  		
$\psi_i$ represents the $i$th Morlet wavelet and 
$\overline{s}$ is the moving average of the signal over a period of $\tau_P=\SI{0.1}{\second}$.
For $i=\{1,...,5\}$, we set $f_{c_i}=\{100, 200, 250, 320, 400\}\si{\hertz}$. The corresponding Strouhal numbers are $St_{H_{c_i}}=f_{c_i} H/U_\infty =\{2, 4, 5, 6.5, 8\}$. Figure \ref{fig:MF1_5} represents the five wavelets in the time and frequency domains. 
\graphicspath{{./Figures_pdf/}}
One may notice that the center frequencies in the frequency domain are slightly different to the values of $f_{c_i}$. This is related to the frequency resolution of the MF which is determined by the wavelet length $\tau_P$ considered in \eqref{eq:morlet_wavelet1}. In the present study, the wavelet includes $200$ points for a time window of $\tau_P=\SI{0.1}{\second}$ within the frequency $f_{RT}=\SI{2}{\kilo\hertz}$. This leads to a frequency resolution of about $\Delta f=\SI{10}{\hertz}$ ($\Delta St_{H}=0.2$). The spectra can then be shifted within $\Delta St_{H}=0.2$ with respect to the set ones. 
\end{document}